\documentclass[iop]{emulateapj}

\usepackage{aas_macros}

\usepackage{color}
\usepackage[dvipsnames,svgnames]{xcolor}

\usepackage{natbib}

\usepackage{amsmath}
\usepackage{xspace}
\usepackage{lineno}

\mathchardef\mhyphen="2D

\newlength{\dhatheight}

\newcommand{\code}[1]{\texttt{#1}\xspace}

\newcommand{\unit}[1]{\ensuremath{\mathrm{\,#1}}\xspace}

\newcommand{\msun}{\unit{M_\odot}}

\providecommand\physrep{\ref@jnl{Phys.~Rep.}}%
\providecommand\apjs{\ref@jnl{ApJS}}%
\providecommand{\jcap}{\ref@jnl{JCAP}}%

\newcommand{\feh}         {\mbox{[Fe/H]}}

\newcommand{\masyr}       {mas~yr$^{-1}$}
\newcommand{\mua}         {\mu_{\alpha}\cos{\delta}}
\newcommand{\mud}         {\mu_{\delta}}
\newcommand{\hst}         {\emph{HST}}

\def\spose#1{\hbox to 0pt{#1\hss}}
\def\lta{\mathrel{\spose{\lower 3pt\hbox{$\mathchar"218$}}
     \raise 2.0pt\hbox{$\mathchar"13C$}}}
\def\gta{\mathrel{\spose{\lower 3pt\hbox{$\mathchar"218$}}
    \raise 2.0pt\hbox{$\mathchar"13E$}}}

\shorttitle{}
\shortauthors{Simon et al.}
\shorttitle{Eri~II: A Fossil from Reionization}
\slugcomment{ApJ, in press}

\begin{document}

\title{Eridanus II: A Fossil from Reionization with an Off-Center Star
Cluster}

\author{Joshua D. Simon\altaffilmark{1}, Thomas M. Brown\altaffilmark{2}, Alex Drlica-Wagner\altaffilmark{3,4,5}, Ting S. Li\altaffilmark{1,6,$\ddagger$}, Roberto J. Avila\altaffilmark{2}, Keith Bechtol\altaffilmark{7,8}, Gisella Clementini\altaffilmark{9}, Denija Crnojevi{\'c}\altaffilmark{10}, Alessia Garofalo\altaffilmark{9}, Marla Geha\altaffilmark{11}, David J. Sand\altaffilmark{12}, Jay Strader\altaffilmark{13}, and Beth Willman\altaffilmark{14} }

\affil{$^{1}$ Observatories of the Carnegie Institution for Science, 813 Santa Barbara St., Pasadena, CA 91101, USA}
\affil{$^{2}$ Space Telescope Science Institute, 3700 San Martin
  Drive, Baltimore, MD  21218, USA}
\affil{$^{3}$ Fermi National Accelerator Laboratory, P.O. Box 500, Batavia, IL 60510, USA}
\affil{$^{4}$ Kavli Institute for Cosmological Physics, University of Chicago, Chicago, IL 60637, USA}
\affil{$^{5}$ Department of Astronomy and Astrophysics, University of Chicago, Chicago, IL 60637, USA}
\affil{$^{6}$ Department of Astrophysical Sciences, Princeton University, Princeton, NJ 08544, USA}
\affil{$^{\ddagger}$ NHFP Einstein Fellow}
\affil{$^{7}$ Physics Department, 2320 Chamberlin Hall, University of Wisconsin-Madison, 1150 University Avenue Madison, WI 53706, USA}
\affil{$^{8}$ LSST, 933 North Cherry Avenue, Tucson, AZ 85721, USA}
\affil{$^{9}$ INAF – Osservatorio di Astrofisica e Scienza dello Spazio di Bologna, Via Piero Gobetti 93/3, 40129 Bologna, Italy}
\affil{$^{10}$ University of Tampa, 401 West Kennedy Boulevard, Tampa, FL 33606, USA}
\affil{$^{11}$ Yale University, Department of Astronomy, 52 Hillhouse Avenue, New Haven, CT 06511, USA}
\affil{$^{12}$ Department of Astronomy/Steward Observatory, 933 North Cherry Avenue, Rm. N204, Tucson, AZ 85721, USA}
\affil{$^{13}$ Department of Physics and Astronomy, Michigan State University, East Lansing, MI 48824, USA}
\affil{$^{14}$ NSF’s National Optical-Infrared Astronomy Research Laboratory, 950 N Cherry Avenue, Tucson, AZ 85721, USA}

\begin{abstract}
  We present deep \emph{Hubble Space Telescope} (\emph{HST})
  photometry of the ultra-faint dwarf galaxy Eridanus~II (Eri~II).
  Eri~II, which has an absolute magnitude of $M_{V} = -7.1$, is
  located at a distance of 339~kpc, just beyond the virial radius of
  the Milky Way.  We determine the star formation history of Eri~II
  and measure the structure of the galaxy and its star cluster.  We
  find that a star formation history consisting of two bursts,
  constrained to match the spectroscopic metallicity distribution of
  the galaxy, accurately describes the Eri~II stellar population.  The
  best-fit model implies a rapid truncation of star formation at early
  times, with $>80\%$ of the stellar mass in place before $z \sim 6$.
  A small fraction of the stars could be as young as 8~Gyr, but this
  population is not statistically significant; Monte Carlo simulations
  recover a component younger than 9~Gyr only 15\% of the time, where
  they represent an average of 7$\pm$4\% of the population.  These
  results are consistent with theoretical expectations for quenching
  by reionization.  The \emph{HST} depth and angular resolution enable
  us to show that Eri~II's cluster is offset from the center of the
  galaxy by a projected distance of $23 \pm 3$~pc.  This offset could
  be an indication of a small ($\sim50-75$~pc) dark matter core in
  Eri~II.  Moreover, we demonstrate that the cluster has a high
  ellipticity of $0.31^{+0.05}_{-0.06}$ and is aligned with the
  orientation of Eri~II within $3 \pm 6$ degrees, likely due to tides.
  The stellar population of the cluster is indistinguishable from that
  of Eri~II itself.
\end{abstract}


\section{INTRODUCTION}
\label{intro}

Nearly 50 years ago, \citet{einasto74} pointed out a striking trend
among the satellites of the Milky Way and other large spiral galaxies:
the dwarf galaxies located closest to a massive galaxy are almost
exclusively dwarf spheroidals containing little or no gas, while more
distant satellites are largely dwarf irregulars hosting significant
amounts of neutral gas.  \citet{einasto74} and subsequent authors
\citep[e.g.,][]{lf83,vdb99} attributed this spatial segregation of
different morphological types to ram-pressure stripping by hot halo
gas.  \citet{br00} brought this result into sharper focus by showing
that around the Milky Way and M31, gas-rich dwarfs are strictly
limited to distances greater than 250~kpc.\footnote{\citet{br00}
  ignored the potential counterexample of the Magellanic Clouds
  because they were focused on much lower-mass satellites that can be
  stripped more easily.  Of course, we now know that the explanation
  for the presence of the gas-rich Magellanic Clouds so close to the
  Galaxy is that they were first accreted by the Milky Way quite
  recently \citep{besla07,kallivayalil13}.}  The discovery of
plentiful star-forming dwarf galaxies around Milky Way-mass hosts by
\citet{geha17} suggests that satellites more luminous than the Milky
Way dwarf spheroidals ($M_{V} \lesssim -13$) may not obey this rule as
they are more difficult to strip.

The discovery of large numbers of dwarf galaxies in the Local
Group over the past 15 years
\citep[e.g.,][]{willman05a,zucker06b,martin06,belokurov07,mcconnachie08,richardson11,bechtol15,koposov15,drlica15}
has not changed this picture.  With the single exception of Leo~T
\citep{irwin07,ryanweber08}, located at a distance of 409~kpc
\citep{clementini12}, all of the recently-identified Milky Way and M31
satellites are devoid of gas
\citep[e.g.,][]{bf07,gp09,spekkens14,westmeier15,mp18}.

A particularly interesting object in this regard is Eridanus~II
(Eri~II), discovered by \citet{bechtol15} and \citet{koposov15}.  At a
distance of 366~kpc \citep{crnojevic16}, Eri~II joins Leo~T as the
only newly discovered dwarfs that appear to be associated with the
Milky Way but are currently outside its presumed virial radius.  The
two galaxies also have similar luminosities, with Eri~II fainter by a
factor of $\sim2$ \citep{dejong08,crnojevic16}.  However, while Leo~T
contains $4.1 \times 10^{5}$~M$_{\odot}$ of neutral hydrogen
\citep{adams18}, the upper limit on the gas content of Eri~II is more
than two orders of magnitude smaller \citep{crnojevic16}.  On the
other hand, \citet{koposov15} identified a handful of bright blue
stars coincident with Eri~II whose colors and luminosities are
consistent with a 250~Myr stellar population, suggesting that star
formation might have continued until very recently.  Later
spectroscopy of 5 out of the 7 candidate young stars by \citet{li17}
showed that they are not associated with Eri~II, casting significant
doubt on the hypothesis of recent star formation.  These observational
results raise obvious questions: Did Eri~II lose its gas within the
last few hundred Myr?  Could environmental processes have been
responsible for ending its star formation despite its large distance
from the Milky Way?

Adding to the intrigue surrounding Eri~II is the possible presence of
a central star cluster in the galaxy.  In the discovery images from
the Dark Energy Survey, \citet{koposov15} noted a ``curious fuzzy
object which can be interpreted as a very faint GC [globular
  cluster].''  \citet{crnojevic16} confirmed the identification of
this cluster using deeper ground-based imaging, making Eri~II the
lowest-mass galaxy known to host a cluster by several orders of
magnitude.  At around the same time, \citet{cusano16} detected a
stellar overdensity near the center of the dwarf spheroidal
Andromeda~XXV, which at $\sim10\times$ more luminous than Eri~II would
be the next faintest galaxy containing a cluster.  \citet{brandt16}
showed that the survival of a cluster near the center of a dwarf
galaxy has significant implications for the viability of dark matter
made up of massive compact halo objects \citep[see
  also][]{li17,zoutendijk20}, and \cite{marsh19} pointed out that the
cluster could also be used to constrain models of ultra-light dark
matter.

Motivated by the possible presence of young stars, the lack of gas,
and its distance beyond the apparent quenching radius of the Milky
Way, we obtained deep \emph{Hubble Space Telescope} (\emph{HST})
imaging of Eri~II to determine its star formation history (SFH).
These data reach below the main sequence turnoff of Eri~II for the
first time.  In this paper we report our measurements of the SFH and
structure of Eri~II using the \emph{HST} photometry.  In
Section~\ref{sec:obs} we describe our processing of the data and
photometric procedures.  We determine the distance of Eri~II and
derive its SFH in Section~\ref{sec:sfh}.  We measure the structural
parameters of Eri~II and its cluster in Section~\ref{sec:structure}.
In Section~\ref{sec:discussion} we discuss the implications of our
results for the evolution of the galaxy and the cluster, as well as
various dark matter models, and in Section~\ref{sec:conclusion} we
summarize our results and conclude.

\section{HUBBLE SPACE TELESCOPE DATA}
\label{sec:obs}

\subsection{Observations}

We observed Eri~II with the Wide Field Channel of the Advanced Camera
for Surveys (ACS; \citealt{acs}) on \emph{HST} through program
GO-14234 (PI: Simon).  The observations were scheduled over seven
visits between 2016 January 16 and 2016 February 8.  We devoted four
visits (8 orbits) to imaging in the F814W filter, totaling 20680~s.
The remaining three visits (two visits of 2 orbits each and one single
orbit visit) were used to image Eri~II in the F606W filter, totaling
12830~s.  The exposure times were chosen in order to reach a
signal-to-noise ratio (S/N) of 10 at 1~mag below the oldest main
sequence turnoff in each filter.  This target S/N is significantly
lower than that achieved by the ultra-faint dwarf imaging by
\citet{brown12,brown14} because Eri~II is $\sim3-9\times$ farther away
than the previously observed systems, but it is still sufficient to
provide useful constraints on the star formation history
\citep[e.g.,][]{gallart99,monelli10b,hidalgo11,weisz12,weisz14a,geha15,skillman17,albers19}.

\begin{figure*}[t!]
\includegraphics[width=6.5in]{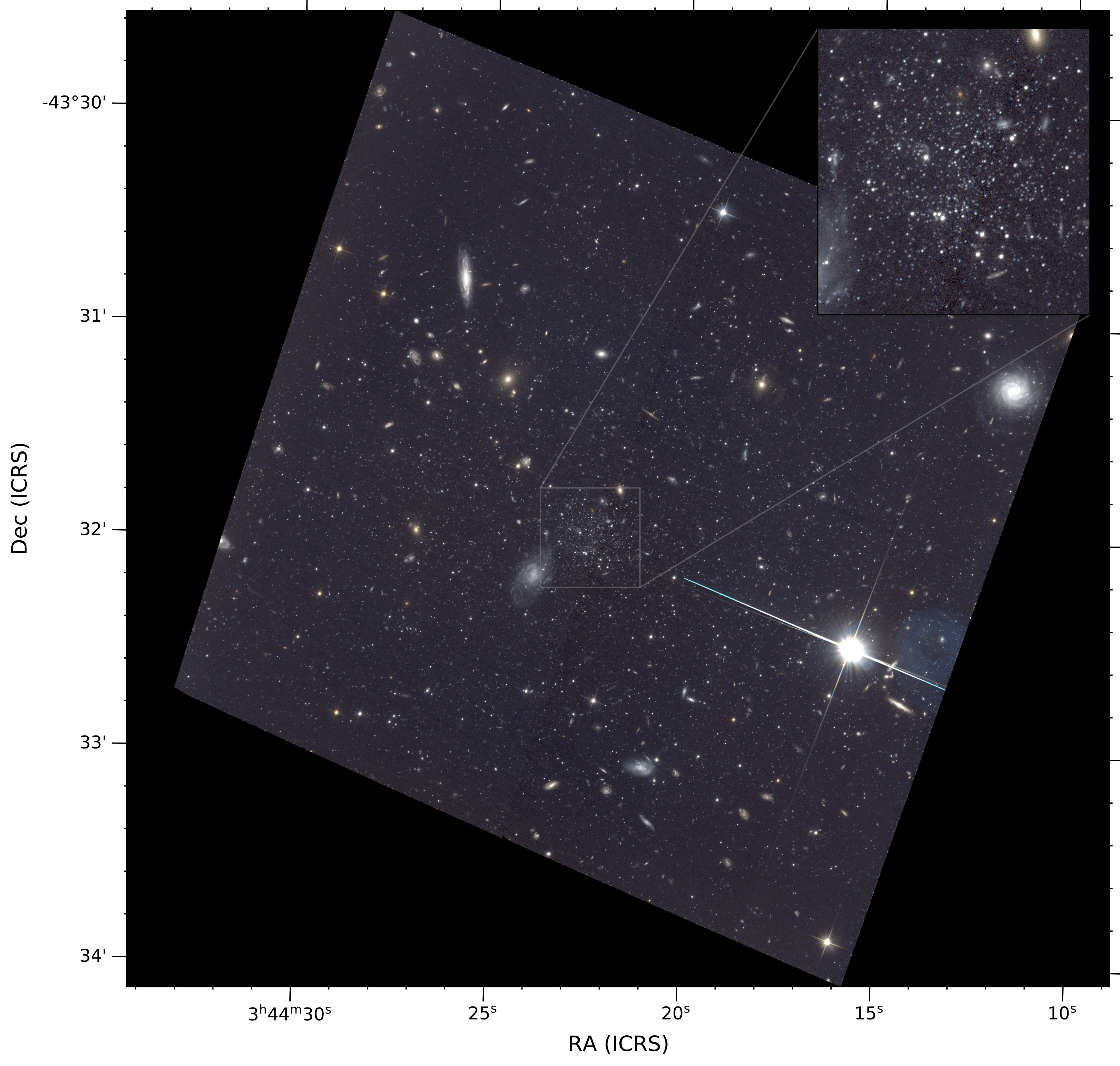}
\caption{ACS color image of Eri~II.  Since we only have data in two
  filters, the F606W image is used for the blue channel, the F814W
  image is used for the red channel, and the green channel is the
  average of the two.  The ACS field of view subtends approximately
  the half-light radius of Eri~II, so the galaxy itself is not
  visually obvious in the image even though essentially all stars
  detected in the field are Eri~II members.  The cluster is visible as
  a much denser concentration of stars slightly below and to the left
  of center, and is highlighted in a $28\arcsec \times 28\arcsec$
  inset.  At the distance of Eri~II (Section~\ref{sec:distance}),
  $1\arcmin$ corresponds to 99~pc.}
\label{eri2_img}
\end{figure*}

Our ACS observations cover a single $202\arcsec \times 202\arcsec$
field centered on Eri~II (see Figure~\ref{eri2_img}).  The half-light
radius ($r_{\rm half}$) of Eri~II is $2.31\arcmin \pm 0.12\arcmin$
\citep{crnojevic16}, so the ACS field of view spans somewhat less than
$1~r_{\rm half}$.  All 7 visits were executed at the same orientation
in order to provide uniform depth over the entire field.  Each 2-orbit
visit placed exposures on a $2\times2$ dither pattern to improve the
sampling of the ACS point spread function (PSF), allow mitigation of
detector artifacts, and enable rejection of cosmic rays.  Successive
visits were stepped by 40\%\ of the gap between the ACS detectors for
the F814W images and 120\%\ of the gap for the F606W images.  The
shorter F606W visit employed the first and third steps of the PSF
resampling dither and was positioned between the two larger offsets
spanning the detector gap.

\subsection{Reduction and Photometry}

\begin{figure}[th!]
\epsscale{1.2}
\plotone{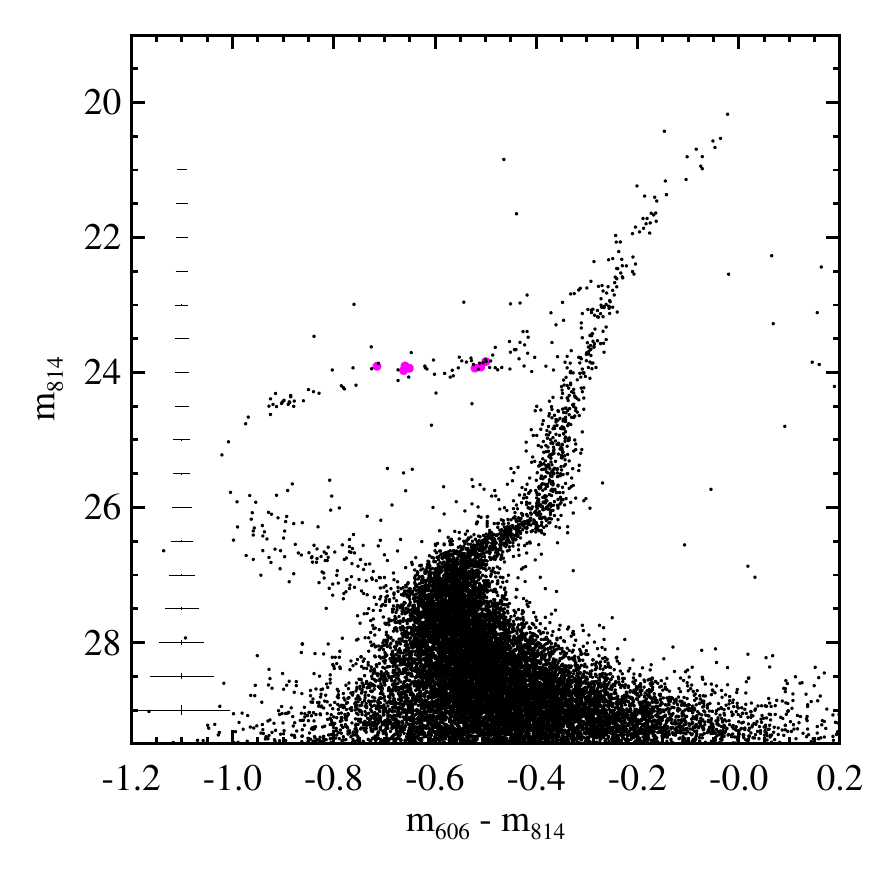}
\caption{Color-magnitude diagram of Eri~II.  Magnitude measurements
  are in the STMAG system, and typical photometric uncertainties as a
  function of magnitude are shown on the left side.  The number of
  Milky Way foreground stars present in the CMD is very low, with
  negligible contamination around the upper main sequence, subgiant
  branch, and lower giant branch that are used for determining the
  star formation history.  The mean colors and magnitudes for seven
  RR~Lyrae stars from Garofalo et al. (in prep.) are plotted as filled
  magenta circles.}
\label{cmd_eri2}
\end{figure}

The process for reducing the data and deriving photometric catalogs
was that used by \citet{brown14}, to which we refer the reader for
details.  In brief, the images were processed with the latest ACS
pipeline, including subtraction of darks and biases, flagging of
detector artifacts, and correction for charge transfer inefficiency.
The individual images for each bandpass were then registered,
resampled, geometrically corrected, and coadded, producing final
images with a scale of 0.\arcsec035/pix and an area of approximately
210\arcsec $\times$ 220\arcsec.  We then performed aperture and
PSF-fitting photometry using the DAOPHOT-II package \citep{stetson87},
producing a catalog in the STMAG system.  The resulting
color-magnitude diagram (CMD) is displayed in Fig.~\ref{cmd_eri2}, and
the measurements for all detected stars are listed in
Table~\ref{photom_table}.  The CMD of Eri~II extends about two
magnitudes below the main sequence (MS) turnoff.  The galaxy is
characterized by a well defined red giant branch (RGB) and a very
extended horizontal branch (HB) that is populated on both sides of and
across the classical instability strip (IS). A prominent blue
straggler sequence extends blueward of the MS turnoff.  Through a
separate reduction of the individual ACS exposures, we have identified
RR~Lyrae stars in Eri II (Garofalo et al., in preparation).  A full
characterization of seven of those stars, including their light
curves, periods, and mean magnitudes, will be presented by Garofalo et
al. (in prep.).

\begin{deluxetable*}{lllllllr}
\tablecaption{Eri~II Stellar Photometry}
\tablewidth{0pt}
\tablehead{
\colhead{Star} & \colhead{RA (J2000)} & \colhead{Dec J2000)} & \colhead{$m_{606}$}
& \colhead{$\delta m_{606}$} & \colhead{$m_{814}$} & \colhead{$\delta m_{814}$}
& \colhead{Flag\tablenotemark{a}}
}
\startdata
    1 &  56.066504 & $-$43.567203 & 28.369 & 0.043 & 29.145 & 0.068 &    0 \\
    2 &  56.067151 & $-$43.566783 & 28.401 & 0.056 & 28.915 & 0.065 &    0 \\
    3 &  56.065480 & $-$43.566400 & 28.727 & 0.080 & 28.888 & 0.059 &   64 \\
    4 &  56.069306 & $-$43.566333 & 29.105 & 0.084 & 29.527 & 0.077 &    0 \\
    5 &  56.066171 & $-$43.566216 & 25.686 & 0.027 & 26.105 & 0.029 &    0 \\
    6 &  56.068574 & $-$43.565958 & 28.427 & 0.044 & 28.858 & 0.056 &    0 \\
    7 &  56.070531 & $-$43.565746 & 28.174 & 0.091 & 28.773 & 0.100 & 1039 \\
    8 &  56.071340 & $-$43.565622 & 29.561 & 0.101 & 29.888 & 0.113 & 1024 \\
    9 &  56.069295 & $-$43.565489 & 27.197 & 0.024 & 27.765 & 0.028 &  512 \\
   10 &  56.068095 & $-$43.565436 & 29.710 & 0.119 & 29.429 & 0.089 &    0 \\
\enddata
\tablenotetext{a}{Flag values are: 1 (fails $\chi$ criterion in F606); 2 (fails $\chi$ criterion in F814); 4 (fails sharp criterion in F606); 8 (fails sharp criterion in F814); 16 (bright neighbor within 4 pixels); 32 (bright neighbor within 8 pixels); 64 (bright neighbor within 12 pixels and 2.5~mag); 128 (bright neighbor within 16 pixels and 2.5~mag); 256 (bright neighbor within 20 pixels and 1.5~mag); 512 (bright neighbor within 24 pixels and 1.5~mag); 1024 (fails photometric uncertainty criterion in both bands).  }
\tablecomments{(This table is available in its entirety in machine-readable form.)}
\label{photom_table}
\end{deluxetable*}

To characterize the photometric uncertainties and completeness as a
function of color and magnitude, over $5\times10^{6}$ artificial stars
were blindly inserted and recovered from the images, adding small
numbers of stars at a time to avoid significantly affecting the
stellar crowding, and including the effects of charge transfer
inefficiency on the recovered S/N.  The 90\%\ completeness limits are
$m_{606} = 28.70$ and $m_{814} = 29.15$, and 50\%\ completeness is
reached at $m_{606} = 29.36$ and $m_{814} = 29.74$.

We used the brightest stars in the field ($m_{606} < 21.5$) to place
the \emph{HST} astrometry in the reference frame of the second data
release (DR2; \citealt{gaiadr2brown,gaiadr2lindegren}) of the
\emph{Gaia} mission \citep{gaia16a}.  Based on the positions of 29
stars with both \emph{HST} and \emph{Gaia} positions, the native
\emph{HST} astrometry was offset from the \emph{Gaia} coordinates by
0\farcs48.  After correcting this offset, the \emph{HST} coordinates
of the bright stars agree with the \emph{Gaia} measurements with a
standard deviation of 0\farcs02.  All coordinates given in this paper
have been shifted to the \emph{Gaia} frame.

\section{THE STAR FORMATION HISTORY OF ERI~II}
\label{sec:sfh}

\subsection{Metallicity Distribution, Distance, and Reddening}
\label{sec:distance}

In order to provide a zero point for comparing theoretical isochrones
to the observed color-magnitude diagram, the metallicity of Eri~II
stars, the distance to the galaxy, and the reddening along the line of
sight must be determined first.  

\begin{figure}[th!]
\epsscale{1.24}
\plotone{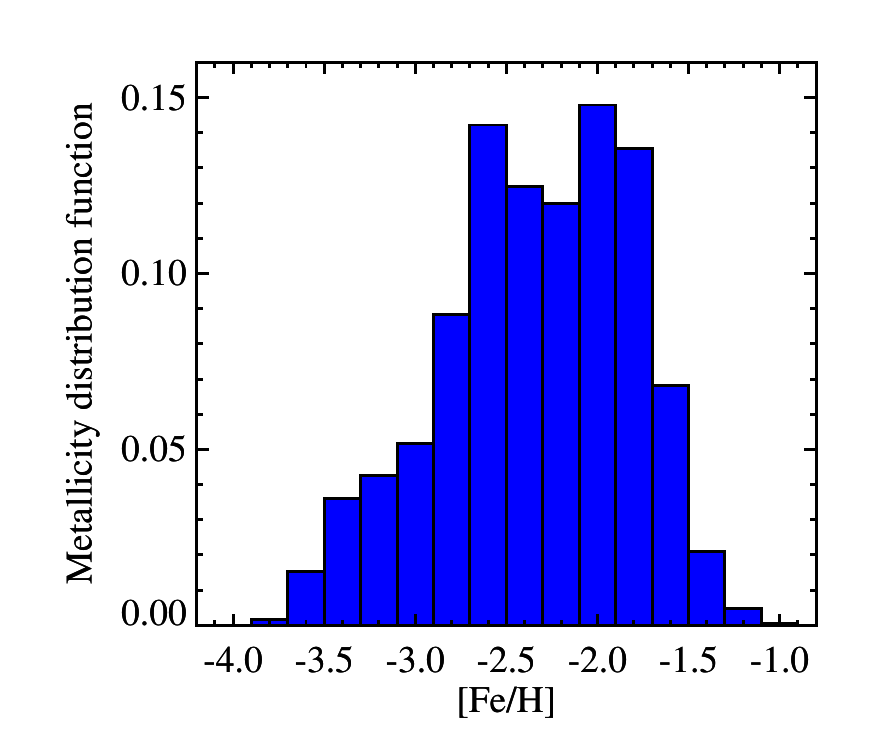}
\caption{Metallicity distribution function of Eri~II, using
  spectroscopic metallicity measurements of 16 stars from \citet{li17}.}
\label{mdf_eri2}
\end{figure}

\citet{li17} measured the metallicities of 16 stars in Eri~II based on
spectroscopy of the Ca triplet absorption lines.  We used those
metallicities to construct a metallicity distribution function (MDF)
by modeling the metallicity of each star as a Gaussian probability
distribution function (PDF), with the Gaussian dispersion set as the
uncertainty of that metallicity measurement.  We then summed the 16
individual PDFs to create a cumulative distribution of Eri~II
metallicities.\footnote{Because the spatial coverage of the
  \citet{li17} spectroscopy is larger than the ACS field of view, 3 of
  the 16 stars with metallicities are located outside the boundary of
  our \emph{HST} imaging.  However, the mean metallicity of the more
  distant stars agrees with that of the sample as a whole, indicating
  that the MDF should not be biased by including them.}  We note that
this method results in an MDF that is artificially broadened relative
to the intrinsic MDF because it effectively double-counts the
measurement uncertainties.  However, the width of the Eri~II MDF is
dominated by real star-to-star metallicity variations rather than
measurement uncertainties \citep{li17}, so this broadening is a minor
effect.  Finally, we made $10^{5}$ draws of 16 samples from the
cumulative distribution to determine the fraction of stars expected in
each 0.2~dex bin between $\feh = -4$ and $\feh = -1$ (see
Figure~\ref{mdf_eri2}).  We adopt this distribution as the Eri~II MDF
for the remainder of the paper.

\begin{figure}[th!]
\epsscale{1.2}
\plotone{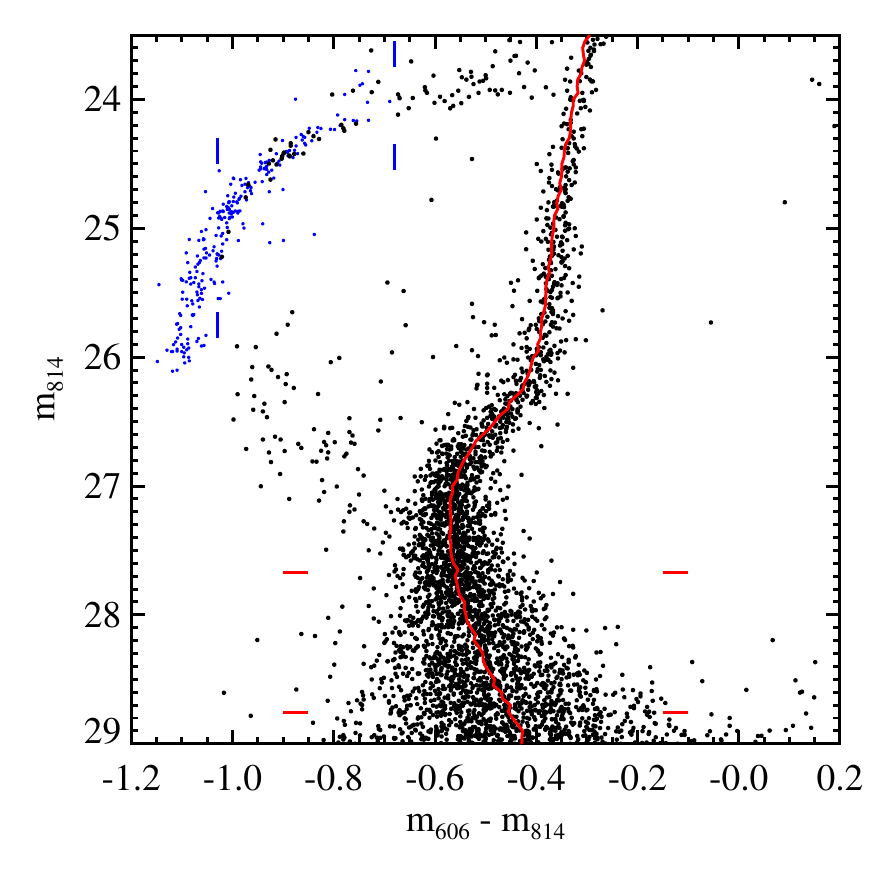}
\caption{Illustration of distance and reddening fit for Eri~II.  The
  black points are the Eri~II photometry, the red curve is the
  ridgeline of the synthetic Victoria-Regina model shifted to the
  measured distance modulus and reddening, and the blue points are
  horizontal branch stars in M92 shifted to the measured distance
  modulus and reddening.  The red and blue tick marks indicate the fit
  ranges for the main sequence and the horizontal branch,
  respectively.}
\label{fig:distance}
\end{figure}

Next, we estimated the distance and reddening.  As in \citet{brown14},
we compared the Eri~II main sequence to a synthetic Victoria-Regina
theoretical isochrone from \citet{vdb14} and the horizontal branch of
Eri~II to the horizontal branch of M92, which was observed with the
same ACS filters by \citet{brown05}.  The synthetic main sequence
isochrone was constructed assuming an age of 13~Gyr, the metallicity
distribution displayed in Figure~\ref{mdf_eri2}, and a binary fraction
of 0.48 \citep{geha13}.  For M92 we assumed a distance modulus of
14.62~mag (the mean of literature measurements by
\citealt{delprincipe05}, \citealt{sollima06}, and \citealt{paust07})
and reddening of $E(B-V) = 0.023$~mag \citep{sfd98}.  We included main
sequence stars more than 0.5~mag below the main sequence turnoff
($m_{814} > 27.67$) to avoid portions that are age-sensitive and we
used horizontal branch stars that are bluer than $m_{606} - m_{814} =
-0.68$ to avoid the RR~Lyrae instability strip.  Fitting the main
sequence and the horizontal branch simultaneously, we found $m-M =
22.65$~mag ($d=339$~kpc) and $E(B-V) = 0.03$~mag, as shown in
Fig.~\ref{fig:distance}.  This distance modulus is consistent within
the uncertainties with the distance derived from the Eri~II RR~Lyrae
stars by Garofalo et al. (in prep.).  The statistical uncertainties on
the distance fit are very small (0.015~mag in distance modulus,
0.003~mag in reddening), and certainly dominated by systematics in the
choice of comparison cluster/isochrone and the magnitude and color
range of stars to include.  By analogy to \citet{brown14} we assume
overall uncertainties of 0.07~mag in $m-M$ and 0.01~mag in $E(B-V)$.
Similar to a few of the ultra-faint dwarfs studied by \citet{brown12}
and \citet{brown14}, the distance modulus determined with this
approach is in modest disagreement with some literature results
(\citealt{crnojevic16} measured $m-M = 22.8 \pm 0.1$~mag and
\citealt{koposov15} estimated $m-M = 22.9$~mag, although
\citealt{bechtol15} found a comparable value of $m-M = 22.6$~mag) and
the derived reddening is larger than the $E(B-V) = 0.009$~mag value
obtained by \citet{sf11} \citep[also see][]{mp19}.  However, for
consistency with previous star formation history analyses, we adopt
the derived values for the remainder of this study.


\subsection{Fitting the Star Formation History}
\label{sec:sfhfit}

Our method for determining the star formation history of Eri~II
follows the procedures described by \citet{brown12,brown14} for
analyzing ACS observations of ultra-faint dwarf galaxies.  Using the
distance modulus and reddening determined in
Section~\ref{sec:distance}, we created a set of synthetic single-age,
single-metallicity stellar populations spanning ages from 8 to
14.5~Gyr and metallicities from $\feh = -4$ to $\feh = -1$.  The
synthetic populations were based on \citet{vdb14} isochrones with
$[\alpha/{\rm Fe}] = +0.4$ and the metallicity-dependent [O/Fe]
abundance used by \citet{brown14}.

We constructed Hess diagrams from both the observed CMD and each of
the synthetic simple stellar populations.  We evaluated the
contamination from foreground Milky Way stars in the ACS photometry by
comparing to a simulation from the Besan{\c c}on model
\citep{besancon}.  Consistent with the appearance of the CMD, the
number of Milky Way stars expected in a single ACS pointing is
extremely small, so we assumed that 0.1\%\ of the stars in the region
of interest do not belong to Eri~II.

We then carried out maximum likelihood fits to compare the synthetic
Hess diagrams to Eri~II.  We included only the portion of the CMD in
the vicinity of the main sequence turnoff, which is the most sensitive
to stellar age.  To match the color and absolute magnitude range used
to determine the star formation histories for other ultra-faint dwarfs
\citep{brown14}, we defined a mask tracing out the stellar locus from
$25.81 \le m_{814} \le 27.51$ and $-0.72 \le m_{606} - m_{814} \le
-0.32$.  Only bins within this mask contributed to the fit.  Because
the S/N at the main sequence turnoff of Eri~II is lower than those
achieved for closer galaxies by \citet{brown14}, we widened the mask
slightly to account for the larger photometric uncertainties.  For
each model of interest, we computed the Poisson equivalent of
$\chi^{2}$ derived by \citet{dolphin02} to evaluate the quality of the
fit.

\begin{figure}[th!]
\epsscale{1.2}
\plotone{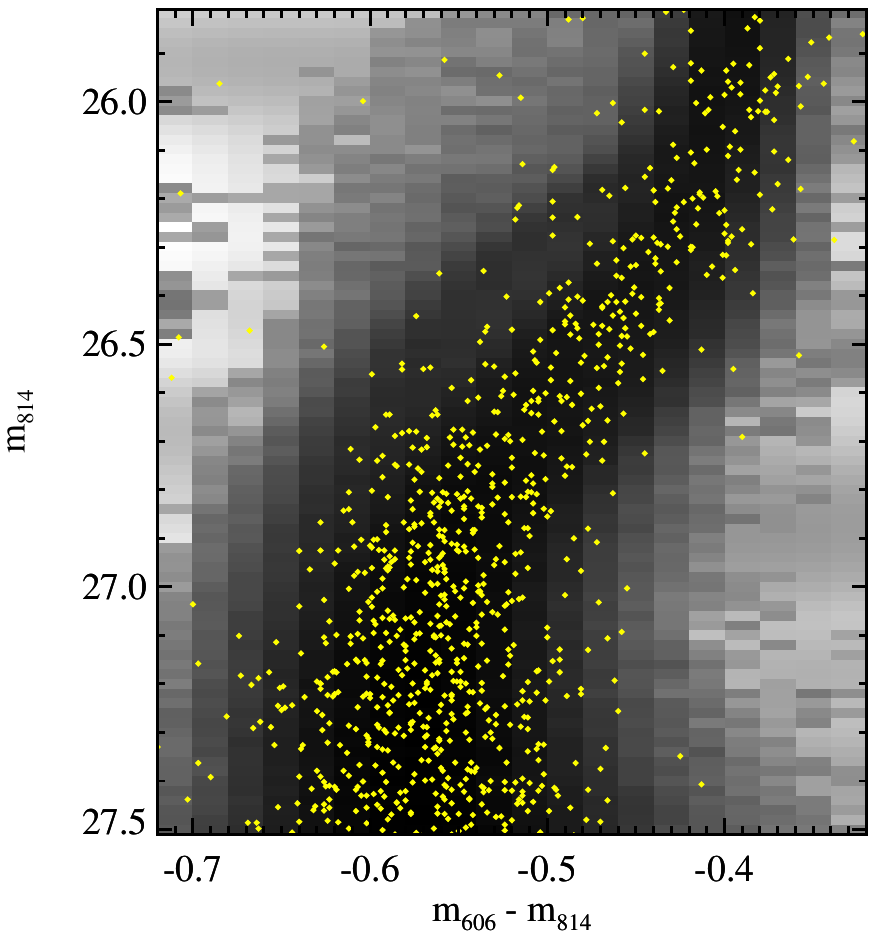}
\caption{Comparison of the best star formation model with the Eri~II
  photometry.  The main sequence turnoff region of the Hess diagram of
  the best two-burst model is shown in gray with a logarithmic
  stretch.  The observed Eri~II stars are plotted as yellow points.}
\label{sfhmodelfig}
\end{figure}

We considered two possible models for the star formation history of
Eri~II.  The simpler one included a single burst of star formation of
negligible duration.  Using a linear combination of theoretical
isochrones to match the observed MDF, the best fit age for the
single-burst model was $12.7^{+0.1}_{-0.6}$~Gyr, with a maximum
likelihood score of 903.  The second model consisted of two bursts of
star formation, plus one additional degree of freedom so that the
older burst is constrained to be more metal-poor than the younger one.
This model still followed the spectroscopic MDF, with all stars below
a threshold metallicity having the older age and more metal-rich stars
forming in the younger burst.  Although this scenario is undoubtedly a
simplification of the actual star formation history, \citet{brown14}
showed that it was a good match to the histories of other ultra-faint
dwarfs.  In the second model, the best fit had 93.7\%\ of the stars in
Eri~II forming 13.5~Gyr ago and 6.3\%\ of the stars being 8.8~Gyr old.
To match the MDF, the division between the older and younger
populations occurs at $\feh = -1.8.$ The mean age for Eri~II in this
model was 13.2~Gyr and the maximum likelihood score was 883.  Because
it offers the best likelihood score and is the most physically
reasonable description of the data, we adopt this three-parameter fit
as the best star formation model for Eri~II.  In
Fig.~\ref{sfhmodelfig} this model is displayed in grayscale, with the
Eri~II stars overlaid.

\begin{figure}[th!]
\epsscale{1.2}
\plotone{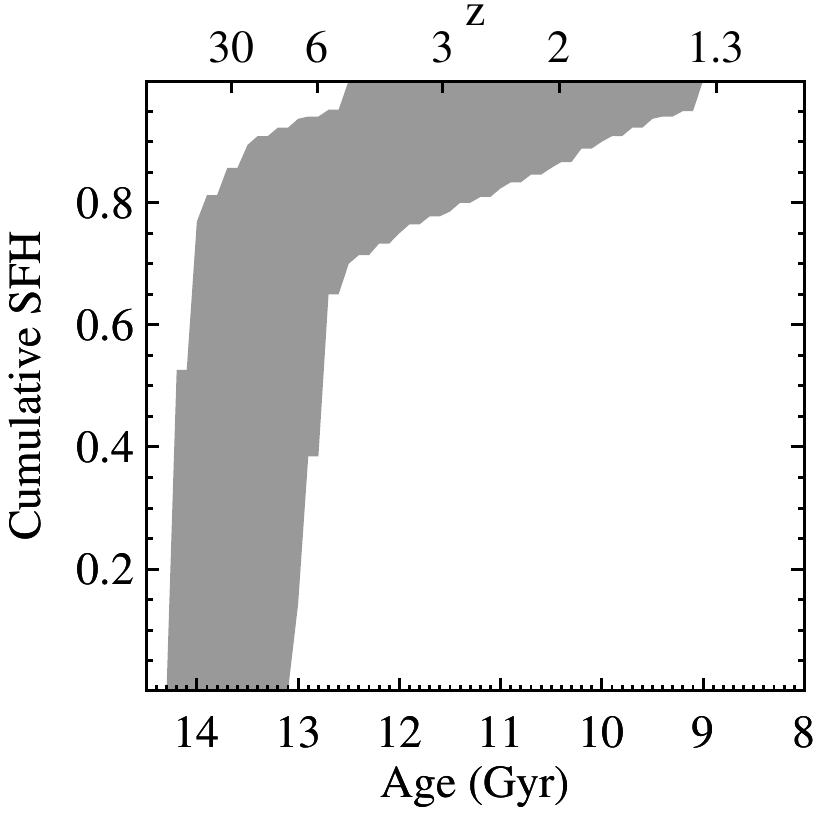}
\caption{Star formation history of Eri~II from a three-parameter fit.
  The bulk of the stars in Eri~II formed before $z=6$ ($\sim13$~Gyr
  ago).  A small amount of additional star formation could have
  extended as long as 4~Gyr, although within the uncertainties the
  star formation may have ended shortly after reionization.}
\label{sfhfig}
\end{figure}

To assess the statistical uncertainties in the star formation history
fits, we created Monte Carlo simulations of the best models.  Using
the best fit model as a seed, we constructed 10,000 mock CMDs with
approximately the same number of stars as the actual Eri~II data set.
We then converted the mock CMDs into Hess diagrams and computed the
maximum likelihood statistic for each mock Hess diagram against the
observed Hess diagram.  For the three-parameter model, the standard
deviation of the maximum likelihood scores from the 10,000 Monte Carlo
iterations was 36, indicating that the difference of 20 (i.e.,
$0.6\sigma$) in the score between a single-burst model and two bursts
for Eri~II is not statistically significant.  Finally, we repeated the
Monte Carlo exercise above, but instead of using the single best model
as the seed for all of the mock CMDs, the seed population was drawn
randomly from all models within one standard deviation of the best
fit.  In Figure~\ref{sfhfig}, we illustrate the range of star
formation histories that are consistent with the data as the shaded
gray band.  We note that the last step of this procedure differs
slightly from the approach of \citet{brown14}, where only the best
model was used as a seed.  Using a broader selection of models in the
Monte Carlo simulation provides a more conservative appraisal of the
uncertainties.

Our results show that $\sim80$\%\ of the stars in Eri~II formed by
$\sim13$~Gyr ago.  In the model consisting of two bursts of star
formation rather than a single burst, the data are consistent with a
small fraction of stars as young as 8--9~Gyr.  However, this possible
younger population is not detected at a statistically significant
level given the uncertainties involved in our star formation history
modeling.  A component younger than 9 Gyr appears in only 15\% of the
Monte Carlo iterations on the fit, and when this component appears, it
accounts for only a small fraction of the total population
($7\pm4$\%).

The discussion above quantifies the statistical uncertainties of our
star formation history fits.  However, as demonstrated by many
previous studies, systematic uncertainties associated with isochrone
libraries are likely to be comparable to or larger than the
statistical uncertainties
\citep[e.g.,][]{dolphin12,weisz14a,skillman17}.  Including sources of
systematic uncertainties such as the chemical composition of Eri~II,
the distance of Eri~II, the distance and reddening of M92, and
following the discussion by \citet{brown14}, the absolute ages we
derive are uncertain at the level of $\sim1$~Gyr.

\subsection{Blue Stragglers}
\label{bs}

As is the case for all old stellar populations, Eri~II contains a
small number of blue straggler stars brighter and bluer than the main
sequence turnoff.  A priori, we do not know whether stars in this part
of the CMD are rejuvenated members of an old stellar population or
represent evidence of more recent star formation
\citep[e.g.,][]{mccrea64,stryker93,bailyn95,ps00}.  To quantify this
population, we followed the methodology established by
\citet{santana13}, who provided a uniform analysis of the blue
straggler abundance across the classical dwarf spheroidals,
ultra-faint dwarfs, and globular clusters.

\begin{figure}[th!]
\epsscale{1.2}
\plotone{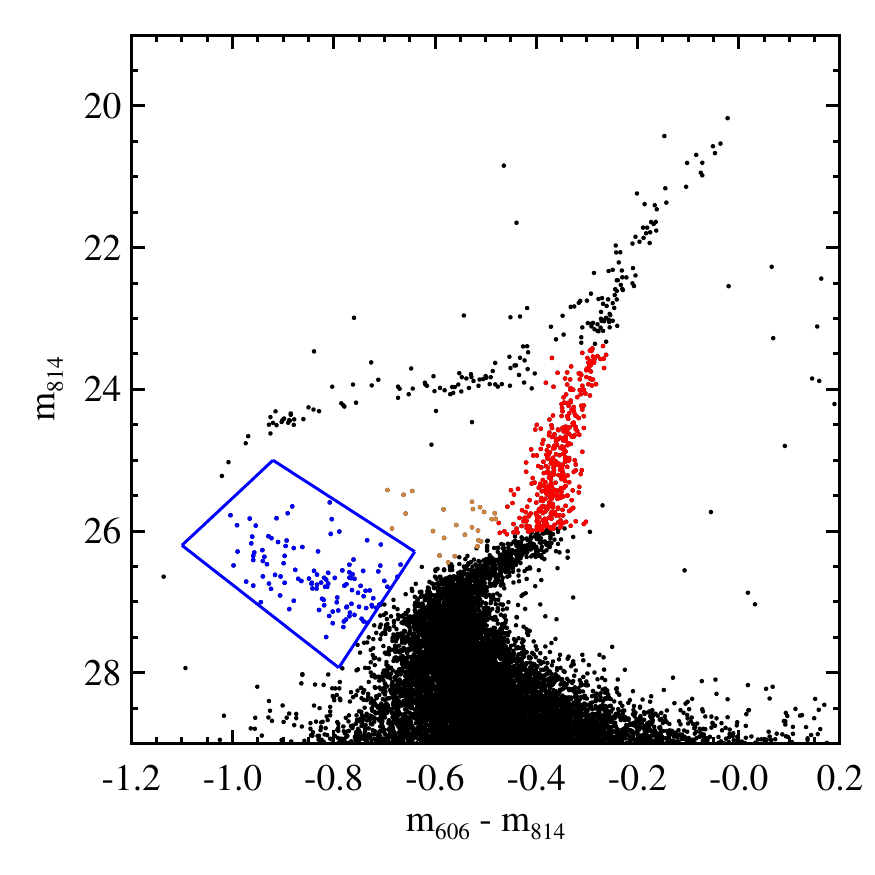}
\caption{Selection of blue stragglers in Eri~II.  The box outlined in
  blue above and blueward of the main sequence turnoff is the
  selection region, and the blue stragglers within the box are plotted
  as blue points.  The stars in the comparison region along the lower
  RGB are plotted in red, and the yellow stragglers are displayed in
  brown. }
\label{cmd_bs}
\end{figure}

\citet{santana13} defined the quantity $F^{\rm BSS}_{\rm RGB}$, which
is the ratio of the number of blue stragglers to the number of stars
on the lower portion of the RGB.  Their color-magnitude selection of
blue stragglers was done empirically for each system, using a box
drawn to maximize the number of blue stars while avoiding
contamination from the MSTO and extreme horizontal branch populations.
To normalize the blue straggler abundance, they used the number of RGB
stars located between 2.4 and 4.9~mag below the tip of the RGB in
$g$-band.  We similarly selected blue stragglers in Eri~II with
customized CMD criteria, as shown in Figure~\ref{cmd_bs}.  The choice
of how closely to allow the blue straggler region to approach the MSTO
is to some degree arbitrary, and the position of the cutoff closest to
the MSTO can change the number of blue stragglers by up to $\sim30$\%.
However, as long as this decision is made in a consistent way for each
system it should not affect the relative values of $F^{\rm BSS}_{\rm
  RGB}$.  
  
Given the difference in filters and photometric systems between this
study and the ground-based $gr$ photometry of \citet{santana13}, the
definition of the RGB selection region does not translate directly.
Instead, we selected RGB stars within a 2.5~mag region (matching the
size of the range used by \citealt{santana13}) extending from the base
of the RGB ($m_{606} = 25.6$) to the top of the horizontal branch
($m_{606} = 23.1$) with a width in color of 0.19~mag, equivalent to
the \citeauthor{santana13} window.  To confirm that these changes do
not affect $F^{\rm BSS}_{\rm RGB}$, we applied the same procedures to
\emph{HST} photometric catalogs for the four galaxies in common
between \citet{santana13} and \citet{brown14}.  In each case, we found
that the $F^{\rm BSS}_{\rm RGB}$ values determined for the \emph{HST}
ACS data sets agree within the uncertainties with the
\citet{santana13} measurements, confirming that the blue straggler
fraction is robust to these choices.

We found $N_{\rm BSS} = 118$ and $N_{\rm RGB} = 391$, yielding $F^{\rm
  BSS}_{\rm RGB} = 0.30 \pm 0.03$.  This value is in perfect agreement
with the result of \citet{santana13} that dwarf galaxies spanning from
the smallest ultra-faint dwarfs to dSphs with $\sim6$ times the
stellar mass of Eri~II share a common value of $F^{\rm BSS}_{\rm RGB}
= 0.29 \pm 0.01$.  The blue straggler population of Eri~II is
therefore entirely consistent with other old stellar populations, and
there is no evidence for an excess of bright main sequence stars
indicating a younger component in Eri~II.  We also examined sources in
the blue straggler region for SX Phoenicis-type variability, with
results presented in Garofalo et al. (in prep.).

In addition to the stars brighter and bluer than the MSTO, Eri~II also
contains a small number of stars above the subgiant branch, between
the blue straggler sequence and the red giant branch (RGB).
Equivalent populations may be present in other dwarfs as well, but are
difficult to assess quantitatively because of significant
contamination from foreground stars and background galaxies (see,
e.g., the CMDs in \citealt{santana13}).  For Eri~II, the combination
of its high Galactic latitude, the small ACS field of view, and the
excellent \hst\ angular resolution, result in negligible contamination
so that these yellow stragglers \citep{hesser84,pz97} stand out
clearly.  We counted $\sim26$ stars in this portion of the CMD.  As
with the blue stragglers themselves, in principle these stars could
indicate a small fraction of younger stars in Eri~II, so we would like
to assess whether there are more of them than would be expected from
the evolution of the blue stragglers as they leave the main sequence.
However, the high incidence of binary stars in this part of the CMD
\citep[e.g.,][]{rozyczka12,kaluzny13,salessilva14,leiner16} and the
complexities of binary evolution make this a difficult task that is
beyond the scope of the present study.  As an alternative, we compared
the ratio of yellow stragglers to blue stragglers to that seen in the
old, metal-poor globular cluster M92 using \hst\ photometry from
\citet{brown05}.  Using the same blue straggler/yellow straggler
definitions as applied to the Eri~II CMD (adjusted for the different
distance modulus and reddening), we found that M92 contains 0.37
yellow stragglers for each blue straggler.  If the same ratio applied
to Eri~II, we would expect to observe $\sim44$ yellow stragglers, well
above the actual number.  This discrepancy may reflect the difference
in blue straggler populations between globular clusters and dwarf
galaxies, but we conclude that the yellow stragglers in Eri~II can
plausibly be explained by evolution of its blue straggler population,
without need for a contribution from younger stars.

\subsection{Comparison of Eri~II and Cluster Stellar Populations}

\begin{figure*}[th!]
\epsscale{1.16}
\plotone{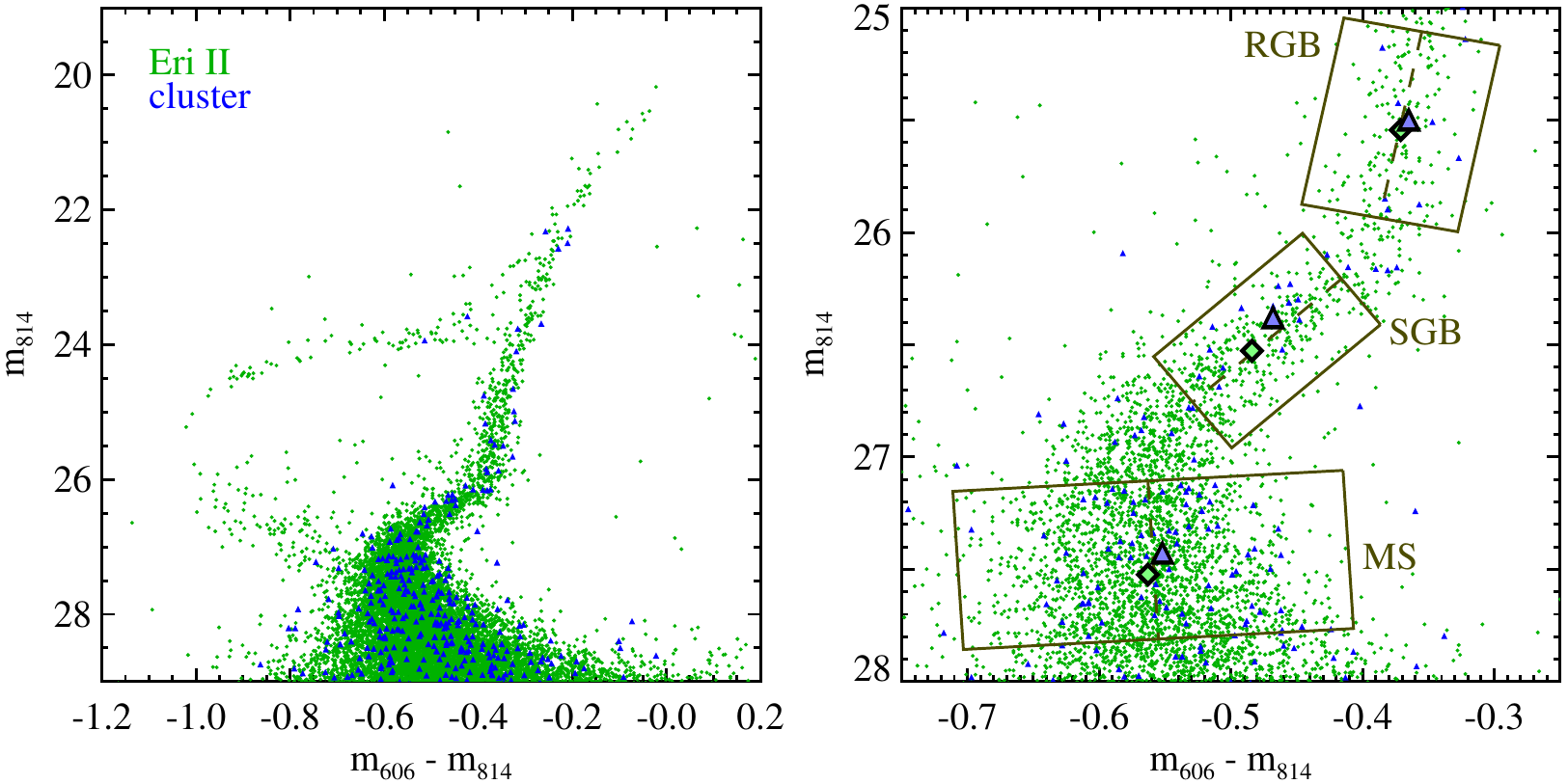}
\caption{Color-magnitude diagram of the star cluster in Eri~II.  The
  blue triangles are stars inside the half-light radius of the
  cluster, and the small green dots, representing Eri~II, are stars
  outside 3 times the cluster's half-light radius.  As described in
  Section~\ref{sec:structure}, the bright neighbor photometry flags
  were ignored in selecting cluster stars because of the crowding near
  the center of the cluster.  In the right panel, the three brown
  boxes centered on the Eri~II ridge line show selection regions for
  (from top to bottom) the lower RGB, the subgiant branch, and the
  main sequence.  Within each box, the large filled green diamond
  indicates the median color and magnitude of the Eri~II stars
  contained in the box, and the large filled blue triangle indicates
  the median color and magnitude for the cluster stars.  The close
  correspondence between the two populations demonstrates that the age
  of the cluster is very similar to that of Eri~II.}
\label{cluster_cmd}
\end{figure*}

Finally, we considered the stellar population of the star cluster in
Eri~II.  As originally argued by \citet{brandt16}, the age of the
cluster affects constraints on dark matter models (also see
Sections~\ref{sec:core} and \ref{sec:machos}).  The CMD of the
cluster, overlaid on that of Eri~II, is displayed in
Fig.~\ref{cluster_cmd}.  To provide maximal separation between the
cluster and Eri~II populations, only stars inside the half-light
radius of the cluster (as determined below in
Section~\ref{sec:structure}) are shown as cluster members here, with
Eri~II members limited to those stars outside 3 cluster half-light
radii.\footnote{The modeling in Section~\ref{sec:structure} shows that
  even in the central region of the cluster, $\sim25\%$ of the stars
  are Eri~II members, so it is not possible to select a pure sample of
  cluster stars.  However, the results below are unlikely to be
  affected by this level of contamination.}  It is visually obvious
that the cluster CMD closely resembles that of Eri~II itself, with the
primary difference being the absence of stars in rare evolutionary
phases (tip of the RGB, blue horizontal branch, and blue stragglers).
We did not detect any RR~Lyrae variables in the cluster.  By drawing
random samples matching the number of cluster stars from the Eri~II
population, we concluded that these differences are not statistically
significant.

As shown by \citet{vdb90}, the color difference between the main
sequence and the base of the RGB is an indicator of relative age when
comparing two populations.  Importantly, this color difference does
not depend significantly on metallicity.\footnote{Eri~II has a mean
  metallicity of $\feh = -2.38 \pm 0.13$, with an internal dispersion
  of $0.47^{+0.12}_{-0.09}$~dex \citep{li17}.  The cluster does not
  have a published metallicity, but \citet{zoutendijk20} measured the
  metallicities of 7 likely cluster members.
  \citeauthor{zoutendijk20} used two different methods to determine
  metallicities, \code{spexxy} \citep{husser12} and The Cannon
  \citep{ness15}.  Based on a comparison with \citet{li17} for one
  star included in both data sets, and for the mean metallicity of
  Eri~II as a whole, we found that the \code{spexxy} metallicities
  appear to be overestimated by $\sim0.3$~dex.  We therefore preferred
  The Cannon metallicities, from which we estimated that the mean
  metallicity of the cluster is $\feh \approx -2$.}

We measured color differences between the main sequence and RGB of
0.192~mag for Eri~II and 0.187~mag for the cluster, so the difference
between Eri~II and the cluster in this quantity is 0.005~mag.  For
comparison, using the \citet{vdb14} isochrones described in
Section~\ref{sec:sfhfit}, at $\feh=-2.4$ (approximately the mean of
the Eri~II MDF), an age increase of 1~Gyr corresponds to a
$(m_{606}-m_{814})_{\mathrm{RGB}} - (m_{606}-m_{814})_{\mathrm{MS}}$
color decrease of 0.009~mag.  By this metric, the cluster is therefore
nominally 0.6~Gyr older than Eri~II, but because of the small number
of stars in the cluster, this age difference is not statistically
significant.

This data set also provided the first opportunity to assess the
distance to the cluster.  Unfortunately, the cluster does not contain
any blue horizontal branch stars, which provided critical leverage on
the distance of Eri~II in Section~\ref{sec:distance}.  Since the
metallicity and age of the cluster are similar to those of Eri~II, as
described above, the luminosity of the subgiant branch can also serve
as a relative distance indicator.  We found a median luminosity
difference in the subgiant box between the cluster and Eri~II of
0.14~mag.  However, as is evident in the right panel of
Fig.~\ref{cluster_cmd}, the small number of subgiants in the cluster
is biased toward the brighter, redder end of the subgiant branch.
Relative to the midline of the subgiant branch box, the difference
between the median luminosity of the cluster stars and the Eri~II
stars is negligible.  We therefore concluded that the cluster distance
is identical to that of Eri~II, with an uncertainty of 0.14~mag in
distance modulus.

\section{THE STRUCTURE OF ERI~II AND ITS CLUSTER}
\label{sec:structure}

In addition to determining the star formation history of Eri~II, our
\hst\ imaging also provides a well resolved look at its central star
cluster for the first time.  The ACS photometry is $\sim3$~mag deeper
than the ground-based data obtained by \citet{crnojevic16}, and the
angular resolution is improved by a factor of $\sim6$.  We therefore
took advantage of this data set to model the spatial structure of both
Eri~II and its star cluster.

We followed the methodology described by \citet{Drlica20} to perform
binned Poisson maximum likelihood fits to the cluster and Eri~II,
modeling each with elliptical profiles as defined by \citet{martin08}.
The stellar catalog used for these fits was limited to magnitudes
where the completeness determined from our artificial star tests was
at least 90\%\ ($m_{606} < 28.70$ and $m_{814} < 29.15$), with an
additional color cut of $m_{606} - m_{814} < 0$ applied to eliminate a
handful of foreground stars.  We also excluded stars flagged as bad by
DAOPHOT, with the exception of those flagged because of bright
neighbors.  A large majority of the stars in the cluster have
neighbors within the DAOPHOT fitting radius, so it is not possible to
accurately model the stellar distribution in the cluster without
including those stars.  Even if the fluxes of the crowded stars are
biased by the neighboring objects (which does not appear to be the
case from the CMD), their positions should be unaffected at any level
that is relevant to determining the size and shape of the cluster.

We determined the best-fitting parameters for Eri II and the cluster
using the affine-invariant Markov chain Monte Carlo ensemble sampler
\code{emcee} \citep{fm13}.  We divided the ACS field of view into $30
\times 30$~pixel ($1\farcs05 \times 1\farcs05$) bins and counted the
number of stars in each bin.  We excluded bins located in the gap
between ACS chips and in the core and diffraction spikes of the
saturated star Gaia~DR2~4836638439545141504 to the west of Eri~II.
Based on tests using mock catalogs generated with various assumptions,
we concluded that we could accurately recover the input parameters
with this procedure and that the bin size does not affect the results.
We first fit Eri~II and the cluster one at a time with the parameters
of the other held fixed to determine the most appropriate functional
form for each system, along with approximate best fit parameter
values.  For the cluster, we attempted fits with exponential,
\citet{plummer11}, \citet{king62}, and \citet{sersic63}
profiles.\footnote{The S{\'e}rsic index and the normalization of a
  S{\'e}rsic profile are tied together by the relation $\Gamma(2n) =
  2\gamma(2n,b_{n})$, where $\Gamma$ is the gamma function, $\gamma$
  is the lower incomplete gamma function, and the surface brightness
  profile is $I(r) = I_{0}e^{-b_{n}(r/r_{h})^{1/n}}$, so it is not
  possible to solve for $b_{n}$ analytically \citep{cb99}.  We
  therefore carried out the S{\'e}rsic fit twice, first with $n$ free
  to vary and the normalization left arbitrary, and then a second time
  with $n$ held fixed at the best-fit value and the normalization
  calculated accordingly.}  The exponential profile overestimated the
central surface density of the cluster, but the other three profiles
all matched the data well.  Any of these profiles would be a
reasonable choice as a model of the cluster, but we identified a
S{\'e}rsic profile with $n=0.41$ as the best fit because it produced
the lowest value of the Akaike information criterion.
\citet{crnojevic16} also determined that a S{\'e}rsic profile was the
best description of the cluster, although their S{\'e}rsic index of
$n=0.19 \pm 0.05$ was somewhat smaller.  For Eri~II itself, we
explored exponential and Plummer fits.  The exponential model is
slightly better over the central $\sim1\arcmin$ of the galaxy, but the
Plummer model is superior in the outer regions.  We decided to move
forward with the Plummer fit for Eri~II based on its improved fit over
most of the imaged area and the better agreement with previous
results.

\begin{figure*}
\epsscale{1.13}
\plottwo{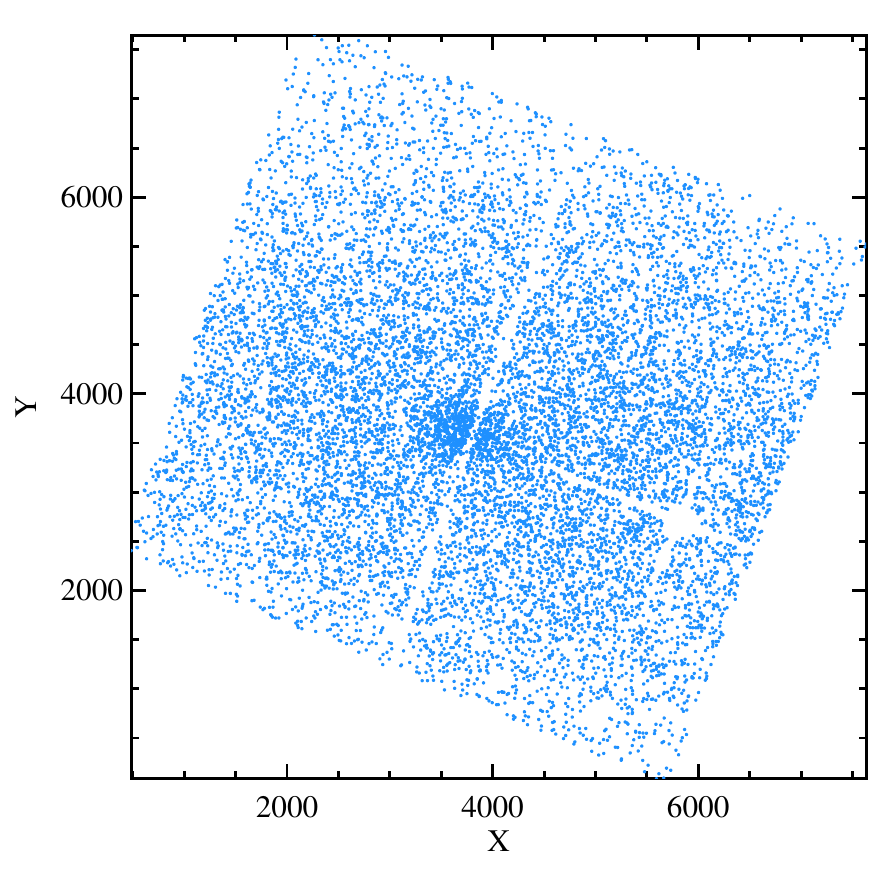}{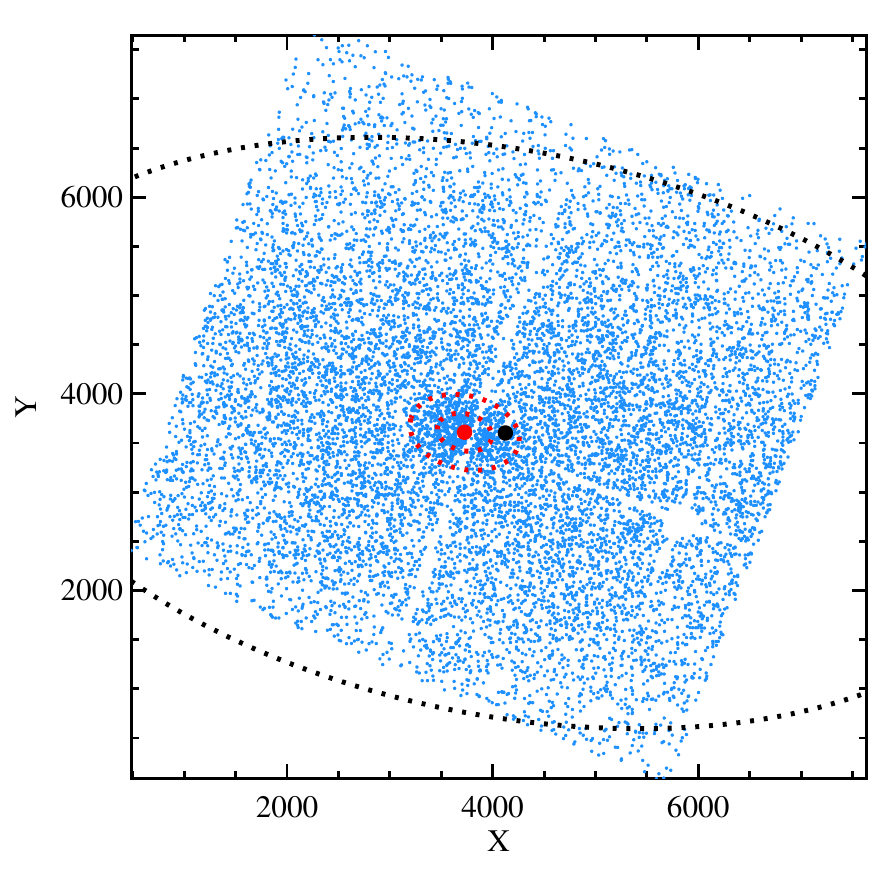}
\caption{Spatial distribution of stars in Eri~II.  The left panel
  shows all stars included in the structural fitting, as described in
  Section~\ref{sec:structure}.  The cluster, ACS chip gap, saturated
  star, and overall orientation of Eri~II are all evident in the
  distribution.  The right panel shows the same set of stars, with the
  central positions of the cluster and Eri~II marked as filled red and
  black circles, respectively.  The uncertainties on each position are
  smaller than the plotted symbols.  The half-light radius of Eri~II
  is displayed as a black dotted ellipse (subject to the caveats
  discussed in the text about the best-fit half-light radius), while
  the red dotted ellipses mark one and two times the half-light radius
  of the cluster.}
\label{spatial_fig}
\end{figure*}

\begin{figure*}
    \centering
    \includegraphics[width=7.0in]{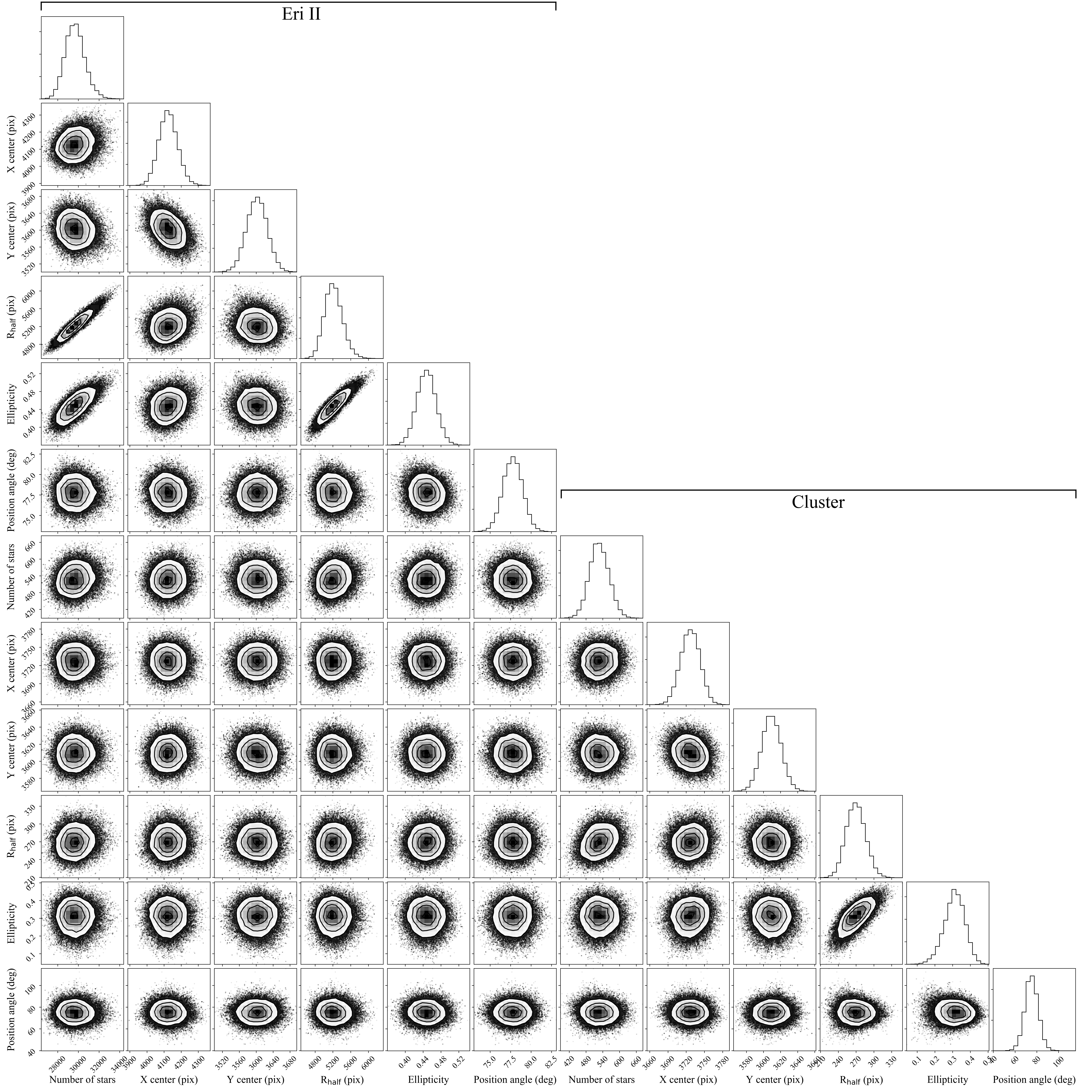}
    \caption{Corner plot for structural fit to Eri~II and its cluster.
      Most of the parameters are uncorrelated with each other, with
      the exception of the ellipticity and half-light radius of each
      component, which exhibit a strong positive correlation, and the
      number of stars in Eri~II, which is also correlated with the
      size and ellipticity of the galaxy.}
    \label{fig:corner}
\end{figure*}

Finally, we modeled both Eri~II and the cluster simultaneously with a
12 parameter fit, using a S{\'e}rsic profile for the cluster and a
Plummer profile for Eri~II.  The free parameters were the center
position, number of stars, half-light radius, ellipticity, and
position angle for Eri~II and the cluster (see
Table~\ref{prior_table}), and the best-fit values are listed in
Table~\ref{struc_table}.  The spatial distribution of stars in the ACS
image and the half-light ellipses of the cluster and Eri~II are shown
in Fig.~\ref{spatial_fig}.  The posterior distributions for each
parameter from the Monte Carlo analysis are displayed in
Fig.~\ref{fig:corner}. For most parameters, the fit appeared to be
well-behaved, with approximately Gaussian posteriors and minimal
correlations between parameters.  However, the half-light radius and
ellipticity for each component were strongly correlated with each
other, in the sense that larger sizes result in higher ellipticities.
The number of stars in Eri~II also exhibited strong positive
correlations with both half-light radius and ellipticity, but the
number of stars in the cluster was only weakly related to the cluster
half-light radius, and not at all to the cluster ellipticity.  The
correlations for Eri~II result from the fact that the imaging only
covers the central portion of the galaxy, so that a more extended
profile requires a larger total number of stars to match the observed
number in the imaged region.

\begin{deluxetable}{llc}
\tablecaption{Prior Parameter Ranges for Structural Fitting}
\tablewidth{0pt}
\tablehead{
\colhead{Row} & \colhead{Parameter} & \colhead{Range} }
\startdata
(1) & Eri~II RA (J2000)             & [03:44:18.3, 03:44:24.8] \\
(2) & Eri~II Dec (J2000)            & [$-$43:32:21, $-$43:31:11] \\
(3) & Eri~II $r_{\text{1/2}}$ (arcsec)  & [70, 280] \\
(4) & Eri~II Ellipticity               & [0.1, 0.9] \\
(5) & Eri~II Position angle (degrees)  & [50, 90] \\
(6) & Eri~II N$_{\text {members}}$       & [5000, 40000] \\
(7) & Cluster RA (J2000)               & [03:44:22.0, 03:44:23.0] \\
(8) & Cluster Dec (J2000)              & [$-$43:32:04, $-$43:31:53] \\
(9) & Cluster $r_{\text{1/2}}$ (arcsec) & [5.25, 28] \\
(10) & Cluster Ellipticity             & [0.0, 0.7] \\
(11) & Cluster Position angle (degrees) & [30, 120] \\
(12) & Cluster N$_{\text{members}}$      & [200, 1200] 
\enddata

\label{prior_table}
\end{deluxetable}

\begin{deluxetable*}{llrr}
\tablecaption{Structural Properties of Eridanus\,II and Cluster}
\tablewidth{0pt}
\tablehead{
\colhead{Row} & \colhead{Quantity} & \colhead{Eri~II} & \colhead{Cluster}
}
\startdata
(1) & RA (J2000)\tablenotemark{a}    & \phs03:44:$21.12 \pm 1\farcs9$  & \phs03:44:$22.40 \pm 0\farcs5$ \\
(2) & Dec (J2000)                          & $-43$:32:$00.1 \pm 0\farcs8$ & $-43$:32:$00.1 \pm 0\farcs4$ \\
(3) & $r_{\rm 1/2}$ (arcsec)\tablenotemark{b} & $182.0 \pm 7.0$ & $9.4 \pm 0.6$ \\
(4) & $r_{\rm 1/2}$ (pc)\tablenotemark{b,c}                    & $299 \pm 12$ & $15 \pm 1$ \\ 
(5) & Ellipticity                          & $0.45 \pm 0.02$ & $0.31^{+0.05}_{-0.06}$ \\
(6) & Position angle (degrees)             & $77.8 \pm 1.2$ & $75 \pm 6$ \\
(7) & S{\'e}rsic index                     &  ... & 0.41 \\
(8) & N$_{\rm members}$\tablenotemark{d}         & $29653^{+968}_{-897}$  & $527^{+36}_{-33}$
\enddata

\tablenotetext{a}{Note that although the R.A. is listed here in hours, minutes, and seconds, the uncertainties are given in arcseconds rather than seconds.}
\tablenotetext{b}{The radii listed here are the semi-major axes of
    the half-light ellipse for each system.  As discussed in the text, the half-light radius of Eri~II is likely overestimated because of the limited spatial coverage of the \emph{HST} imaging.  We include the best-fit value here for completeness, but we expect that the smaller size given by \citet{crnojevic16} is more accurate.}
\tablenotetext{c}{Sizes in physical units are calculated assuming a distance of 339~kpc (\S~\ref{sec:distance}).}
\tablenotetext{d}{The number of member stars (integrated to infinity, not limited to the spatial extent of our imaging) satisfying the constraints $m_{814} < 29.15$, $m_{606} < 28.70$, and $m_{606} - m_{814} < 0$.}
\label{struc_table}
\end{deluxetable*}

\begin{figure*}[th!]
   \centering
    \includegraphics[width=7.0in]{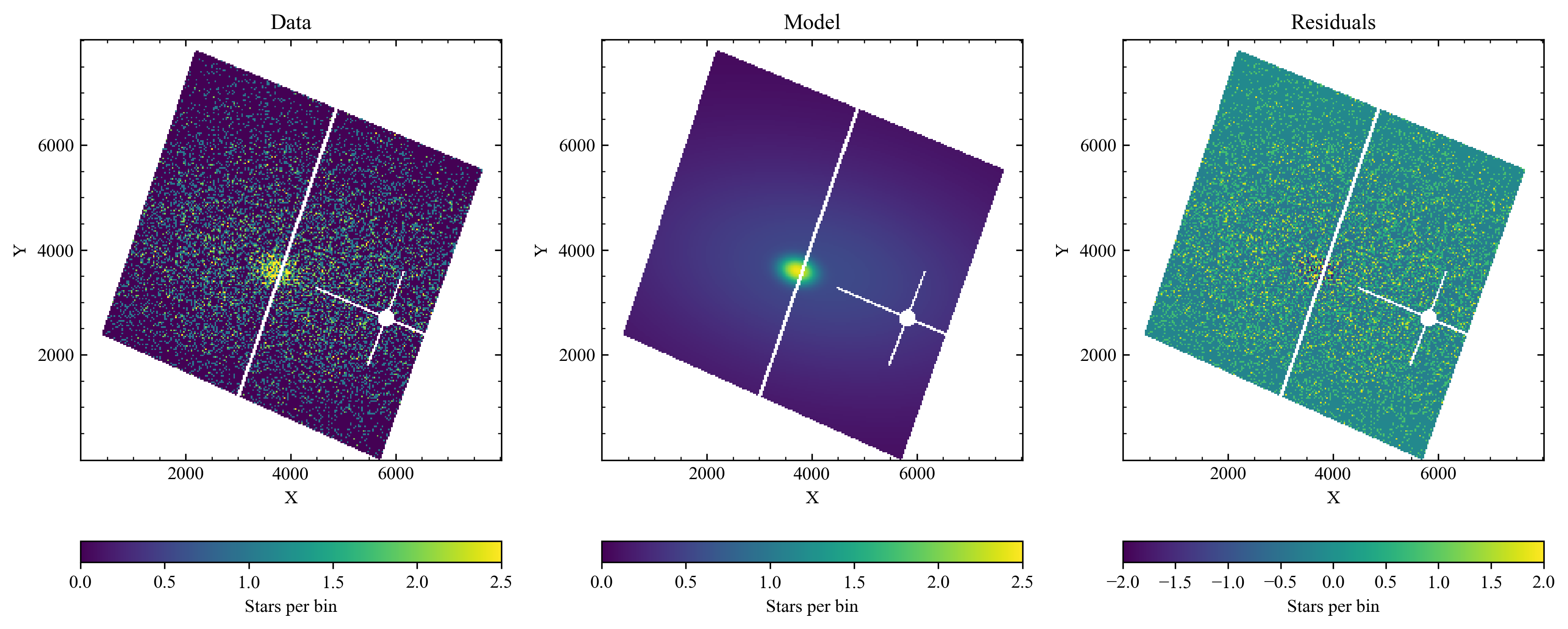}
    \caption{(left) Distribution of stars in the \emph{HST} data in
      $30 \times 30$~pixel bins. White areas were masked out because
      of a lack of data.  (middle) Best-fit model of Eri~II and
      cluster.  (right) Spatial map of residuals after subtracting the
      model from the data, showing a good match between the two.  Note
      the change in scale of the color bar relative to the other two
      panels.}
\label{data_model_residual}
\end{figure*}

The spatially binned and masked data, along with the best-fit model
and the residuals, are illustrated in Fig.~\ref{data_model_residual}.
We found that the cluster is significantly elongated, with an
ellipticity of $e = 0.31^{+0.05}_{-0.06}$.  In earlier ground-based
imaging, the cluster appeared to be round because of the decreased
angular resolution and the small number of resolved stars
\citep{crnojevic16}.  Moreover, the cluster is perfectly aligned with
the orientation of Eri~II: we measured position angles of $77.8 \pm
1.2$ degrees for Eri~II and $75 \pm 6$ degrees for the cluster.  It is
unlikely that this alignment would occur by chance, so this result
points to a common physical mechanism being responsible for the shapes
of the cluster and Eri~II.  We also determined a half-light radius for
the cluster of $9\farcs4$ ($15 \pm 1$~pc), larger than the $6\farcs6
\pm 0\farcs6$ obtained by \citet{crnojevic16}.  Again, the deeper,
higher-resolution \emph{HST} data are likely responsible for this
difference.  Based on the best-fit number of member stars in each
object (integrated out to infinity; \citealt{Drlica20}), the
luminosity of the cluster is $1.8\% \pm 0.1\%$ that of Eri~II, which
translates to $M_{V} = -2.7 \pm 0.3$ for an Eri~II absolute magnitude
of $M_{V} = -7.1 \pm 0.3$ \citep{crnojevic16}.  For comparison,
\citet{crnojevic16} measured an absolute magnitude for the cluster of
$M_{V} = -3.5 \pm 0.6$, somewhat brighter than but roughly consistent
with our result.

We measured central positions of $\alpha = 56.08799$, $\delta =
-43.53335$ for Eri~II and $\alpha = 56.09332$, $\delta = -43.53335$
for the cluster.  The separation between the two is $13\farcs9 \pm
2\farcs0$, or $23 \pm 3$~pc at a distance of 339~kpc.  The cluster is
therefore offset from the center of the galaxy by a small, but
detectable, amount.  \citet{crnojevic16} obtained a slightly larger
offset of 23\arcsec\ (41~pc), but given the uncertainty on the center
of Eri~II in their shallower imaging, this offset was only significant
at the $\sim2\sigma$ level.

Comparing our overall results to those of \citet{crnojevic16}, as
described above, we determined that the central cluster is larger and
more elliptical than previously measured.  For Eri~II, we also found a
larger radius ($182\farcs0 \pm 7\farcs0$) than measured from
ground-based data ($138\farcs6 \pm 7\farcs2$).  Although we were able
to recover the input half-light radius in tests with mock data
designed to match the properties of the \emph{HST} photometric
catalog, we concluded that this discrepancy is likely the result of
the geometry and limited spatial coverage of our imaging.  In
particular, \citet{munoz12} showed that, among other conditions, the
field of view for an imaging program should be at least 3$\times$ the
half-light radius of a system in order to accurately recover its
structural parameters.  Since measuring the size of Eri~II was not one
of the goals of the ACS imaging, the obtained data set is not close to
meeting this criterion.  We therefore suggest that the Eri~II
half-light radius determined by \citet{crnojevic16} from much
wider-field imaging is likely more reliable.  The ellipticity that we
measured for Eri~II agrees with that of \citet{crnojevic16} within the
uncertainties, and the center position and PA are consistent at the
$\sim1-1.5\sigma$ level.

\section{ANALYSIS AND IMPLICATIONS}
\label{sec:discussion}

In this section we consider the implications of our star formation
history and structural measurements for the history of Eri~II and the
properties of dark matter.

\subsection{The Quenching of Eri~II}
\label{sec:quenching}

As mentioned in Section~\ref{intro}, the discovery imaging for Eri~II
contained hints of star formation within the past few hundred million
years \citep{koposov15}.  \citet{li17} cast significant doubt on this
possibility by showing that the brightest stars in this putative young
population are not spectroscopic members of Eri~II.  Our much deeper
imaging analyzed in Section~\ref{sec:sfhfit} confirms that Eri~II is
exclusively an old system, with no sign of stars having formed in the
past $\sim9$~Gyr.  The prevalence of RR~Lyrae variables in Eri~II
\citep[cf.][]{martinezvazquez19,vivas20}, combined with the lack of
anomalous Cepheids (unlike the populations exhibited by Leo~T;
\citealt{clementini12}), supports this conclusion. Here we consider
what these results mean for the quenching of star formation in Eri~II.

\citet{fritz18} used astrometric measurements from the second
\emph{Gaia} data release \citep{gaiadr2brown} to determine the proper
motion of Eri~II.  Based on a sample of 12 spectroscopic member stars
from \citet{li17} that are also in the \emph{Gaia} catalog,
\citet{fritz18} obtained the first proper motion constraint for
Eri~II.  Subsequently, \citet{zoutendijk20} identified additional
Eri~II members via deep VLT/MUSE spectroscopy, but none of those stars
have astrometry in the \emph{Gaia} DR2 catalog.  We adopted the member
sample of \citet{pl19}, which is based on the \citeauthor{li17}
spectroscopic members plus an additional 7 likely member stars
(membership probability $\ge50\%$ and colors consistent with a low
metallicity) identified via photometry and astrometry. We added one
spectroscopically confirmed bright red giant member from \citet{li17}
that is missing from the \citet{pl19} list because its photometry in
the DES DR1 catalog \citep{desdr1} is compromised by its proximity to
the 12th magnitude foreground star visible in Fig.~\ref{eri2_img}.
Using these stars, the weighted average proper motion for Eri~II is
$\mua = 0.25 \pm 0.21$~\masyr, $\mud = 0.03 \pm 0.24$~\masyr.
Unsurprisingly, given the large distance to Eri~II, the proper motion
is consistent with zero within the uncertainties, and all of the
available estimates are mutually consistent.\footnote{\citet{mv20}
  determined a proper motion with much smaller uncertainties based on
  the assumption that Eri~II is gravitationally bound to the Milky
  Way.  Because we are attempting to investigate the orbital history
  of Eri~II, we prefer to avoid making such an assumption.  By
  removing their prior on the tangential velocity, \citet{mv20} find
  $\mua = 0.35^{+0.21}_{-0.20}$~\masyr, $\mud = -0.08 \pm
  0.25$~\masyr, in agreement within the uncertainties with our
  determination.}

\begin{figure*}
\epsscale{1.16}
\plotone{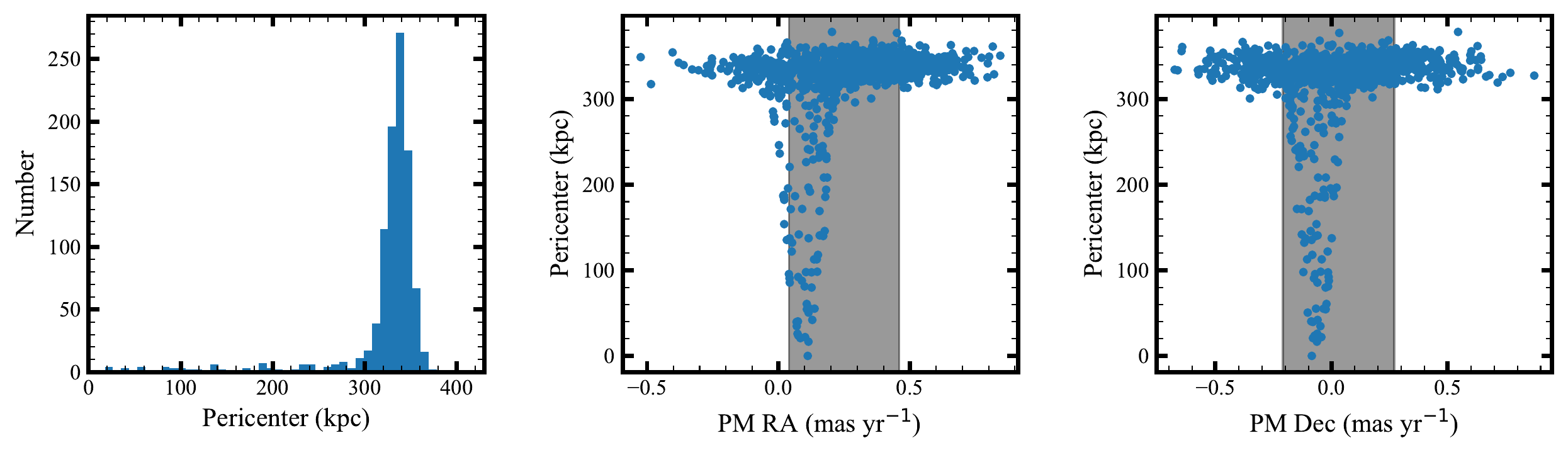}
\caption{Constraints on the orbital pericenter of Eri~II given the
  current observational uncertainties.  The left panel shows a
  histogram of the pericenters calculated with \texttt{galpy} for the
  observed position, distance, velocity, and proper motion of Eri~II.
  The middle and right panels show the pericenter as functions of the
  proper motion in the R.A. and Decl. directions, respectively.  The
  shaded gray bands indicate the $1\sigma$ confidence interval on each
  component of the proper motion.  Pericentric approaches within
  300~kpc of the Milky Way are only possible for a narrow range of
  proper motion values, which can be tested with future measurements.}
\label{fig:pericenter}
\end{figure*}

We used the \texttt{galpy} software developed by \citet{galpy} to
compute the orbit of Eri~II in the Milky Way potential.  We adopted
the modified MWPotential2014 gravitational potential described by
\citet{carlin18}, which has a total mass of $1.6 \times 10^{12}~\msun$
\citep[e.g.,][]{watkins19}.  For the nominal proper motion of Eri~II,
the resulting orbit places the galaxy essentially at the pericenter of
its trajectory around the Milky Way at present.  However, because the
measurement uncertainties are comparable to the proper motion itself,
a substantially broader array of orbits are also consistent with the
data.  To explore the parameter space, we drew 1000 samples of the
radial velocity, proper motion, and distance of Eri~II from Gaussian
distributions of each quantity centered on the measured value and with
the dispersion set equal to the uncertainty.  We then calculated the
orbit for each set of values.  We found a median orbital pericenter of
$335^{+13}_{-18}$~kpc, on a highly eccentric orbit with $e =
0.87^{+0.05}_{-0.17}$.  The orbital properties are independent of the
line-of-sight velocity or distance of Eri~II, but for a narrow range
of proper motions ($\mua \approx 0.10 \pm 0.10$~\masyr, $\mud \approx
-0.05 \pm 0.10$~\masyr) the pericenter can be significantly smaller
($0-300$~kpc).  The observed proper motion is consistent with this
solution (see Fig.~\ref{fig:pericenter}).

Although the proper motion uncertainties for Eri~II will shrink in
future \emph{Gaia} data releases, for the time being the observed
kinematics do not conclusively reveal whether it is a bound satellite
of the Milky Way.  As an alternative avenue for providing insight into
the history of Eri~II, we examined the properties of subhalos drawn
from the ELVIS simulations \citep{gk14} that have similar distances,
stellar masses, and galactocentric velocities as Eri~II, following
\citet{li17}.  Using the full ELVIS data set (including both isolated
and paired Milky Way analogs), the majority of the ELVIS subhalos
compatible with Eri~II have already made at least one pericentric
passage around their host galaxy, with only $\sim22\%$ currently
falling in for the first time.  If future improved proper motion
measurements show that Eri~II is one of the minority of subhalos on
its first infall, then reionization is by far the most likely
explanation for its early quenching.  On the other hand, in the more
common scenario where Eri~II is a backsplash galaxy
\citep[e.g.,][]{warnick08,teyssier12,fillingham18,blana20}, we must
examine the distribution of possible infall times.  Using broader
selection criteria than \citet{li17} to provide a larger sample of
comparison halos, \citet{rw19} showed that Eri~II may have interacted
with the Milky Way at low redshift ($z\lesssim1$), but is unlikely to
have experienced a close passage at earlier times.  We conducted a
similar exercise but with slightly updated distance and velocity
selection cuts, finding that the earliest first infall time among 125
Eri~II analogs in ELVIS is 11~Gyr ago.  On the other hand, most of the
star formation in Eri~II ended at even earlier times, prior to 12~Gyr
ago (see Section~\ref{sec:sfh}).  If the orbital distribution of the
ELVIS subhalos is realistic, we therefore conclude that environmental
influences from the Milky Way are very unlikely to be the dominant
factor controlling star formation in Eri~II.  Instead, as with the
more nearby ultra-faint dwarfs \citep{brown14}, the timing of the
decline in star formation in Eri~II is more consistent with
reionization \citep[e.g.,][]{rg05,br11a,wheeler19}.  However, we
cannot rule out that environmental processes affected low levels of
star formation in Eri~II that may have persisted until somewhat later
times.

\subsection{The Curious Alignment of the Cluster with Eri~II}
\label{sec:tides}

In Section~\ref{sec:structure} we determined that the star cluster is
both significantly elongated, with $e=0.31^{+0.05}_{-0.06}$, and
aligned with the orientation of Eri~II within the measurement
uncertainties, as can be seen visually in Fig.~\ref{spatial_fig}.  The
ellipticity is unusually large for a globular cluster; the highest
ellipticity listed in the \citet[][2010 edition]{Harris96} catalog is
0.27, for NGC~6273 \citep{ws87}.  A handful of faint clusters analyzed
by \citet{munoz18} have comparable shapes, such as Koposov~1 and 2,
AM~4, Mu{\~n}oz~1, Balbinot~1, and Kim~1, although the measurement
uncertainties are generally large.  The size of the cluster is also an
outlier from the Milky Way globular cluster population.  Among
clusters with similar luminosities, only Pal~12, Laevens~1, and DES~1
have $r_{\rm half} \gtrsim 9$~pc.\footnote{More recent measurements of
  Pal~12 by \citet{musella18} indicate a smaller half-light radius of
  5.4~pc and attribute the larger size found previously to
  contamination from the Sagittarius stream.}  Perhaps the closest
known analogs to the Eri~II cluster are Fornax~6 \citep{wang19} and
the candidate cluster in And~XXV, which is somewhat larger and
brighter ($r_{\rm half} = 25$~pc and $M_V = -4.9$;
\citealt{cusano16}).  We note that if the cluster had been discovered
in isolation rather than near the center of a dwarf galaxy, it might
be considered a candidate dwarf galaxy itself, with properties
resembling, e.g., Triangulum~II or Draco~II.  The only obvious way to
test the possibility that the cluster is actually a dwarf would be
with metallicity and chemical abundance measurements of cluster stars
to determine whether it is a monometallic system.  Unfortunately, even
the brightest few stars in the cluster are challenging targets at best
for spectroscopy with current instruments (see Fig.~\ref{cluster_cmd}
and \citealt{zoutendijk20}).

The simplest explanation for the properties of the cluster is that it
is being tidally distorted by the gravitational potential of Eri~II.
To provide an initial assessment of the viability of this hypothesis,
we compute the Jacobi radius of the cluster, as defined by
\citet{bt08}.  Assuming that the velocity dispersion of Eri~II is
constant with radius, as is the case for most Milky Way satellites
\citep[e.g.,][]{walker07,battaglia11,pace20}, the mass profile near
the center of the galaxy is $M(R) = \sigma^{2}R/G$.  Then, for a
cluster mass-to-light ratio of 2~$M_{\odot}/L_{\odot,V}$ and the
estimated absolute magnitude given in Section~\ref{sec:structure}, the
Jacobi radius at the projected radius of the cluster is $\sim3$~pc.
The observed half-light radius of the cluster is 15~pc, indicating
that most of the cluster's stars are currently beyond its tidal
radius.  In order for the Jacobi radius to exceed the half-light
radius, the three-dimensional distance between the cluster and Eri~II
would need to be $\gtrsim220$~pc despite their much smaller projected
separation.  Although this calculation is highly simplified in all
respects, we conclude that it is not surprising that the cluster is
being tidally stretched along the axis connecting the center of the
cluster and the center of Eri~II.

If the instantaneous tidal force of Eri~II is responsible for the
ellipticity of the cluster, the alignment between the two remains to
be explained.  An ellipsoidal or triaxial dark matter halo for Eri~II
with a similar orientation to the stellar component of the galaxy
could have contributed to the shape of the cluster via its
gravitational potential.  On the other hand, one might expect
precession or tumbling motions to decrease such an alignment with
time.  Alternatively, if the cluster moves on a radial orbit that is
aligned with the major axis of Eri~II, then the tidal forces described
above would tend to cause the cluster's elongation to be oriented in
the same direction as well.

A more detailed understanding of the tidal interaction between the
cluster and Eri~II would require tailored N-body simulations
\citep[e.g.,][]{amorisco17,contenta18}, but for now we suggest that
tides appear likely to be responsible for the large size and
ellipticity of the cluster.

\subsection{Does Eri~II Contain a Dark Matter Core?}
\label{sec:core}

\citet{goerdt06} suggested that the positions of globular clusters
within dwarf galaxies can be an indicator of the central dark matter
density profile of the galaxy, potentially providing new insight into
the longstanding cusp-core problem
\citep[e.g.,][]{fp94,moore94,deblok01,kuzio08,relatores19}.  The basic
argument is that in a cuspy dark matter halo, dynamical friction
should rapidly drag clusters to the center of the galaxy, while in a
cored halo the clusters can survive at larger radii for more than a
Hubble time \citep[e.g.,][]{read06,boldrini19,leung20}.  This picture
has primarily been investigated to explain the observational fact that
the globular clusters of the Fornax dwarf spheroidal are located at
typical radii of $\sim1$~kpc.

The discovery of a cluster in the lower luminosity and more dark
matter-dominated galaxy Eri~II offers an even better laboratory for
the exploration of globular cluster dynamics, as already recognized by
\citet{amorisco17} and \citet{contenta18}.  Both
\citeauthor{amorisco17} and \citeauthor{contenta18} showed with N-body
simulations that star clusters in an Eri~II-like galaxy are quickly
tidally disrupted if the galaxy has a cuspy density profile.  Clusters
can only survive in a cuspy galaxy if located exactly at the center of
the halo.  On the other hand, they found that clusters in cored
potentials ($\alpha \lesssim 0.2$, where the inner density profile is
$\rho(r) \propto r^{-\alpha}$) are long-lived and dynamical friction
stalls before the cluster reaches the center of Eri~II.

Unlike in the case of Fornax, the cluster in Eri~II is quite close to
the center of the galaxy.  However, our measurements in
Section~\ref{sec:structure} revealed a small but significant offset of
$23 \pm 3$~pc (in projection) between the two.  Such a small offset is
surprising because the cores often postulated in other dwarfs have
typical sizes of a few hundred pc to $\sim1$~kpc.  Recent calculations
indicate that the radius at which dynamical friction stalls for a
cluster orbiting within a dwarf galaxy is $\sim1/3$ of the radius of
the core \citep{meadows20,kaur18}.  If this mechanism is responsible
for the location of the cluster, the implied core size for Eri~II is
only $\sim70$~pc.  Alternatively, if Eri~II has a cuspy density
profile, clusters with stellar masses similar to the observed value
should be destroyed within a few Gyr at distances of less a few
hundred pc from the center of the galaxy, unless the cluster formed as
a nuclear cluster at the center of the halo \citep{amorisco17}.  Since
a central position for the cluster is disfavored by its apparent tidal
elongation, a cuspy halo can only be consistent with the observed
location in the unlikely scenario that the line-of-sight distance
between Eri~II and the cluster is $\gtrsim1$~kpc, and the cluster is
projected to lie so close to the center by coincidence
\citep{contenta18}.

We note that given the measured stellar mass and size of the cluster,
its relaxation time is less than its age, so two-body effects may be
important to its evolution \citep{contenta18}.  The interaction
between tides and relaxation-driven expansion and mass loss depends on
the mass profile of the galaxy and the position of the cluster
relative to the center of the potential.  A measurement of the
velocity dispersion of the cluster could in principle separate the
cored and cuspy cases, but obtaining accurate enough velocities for
such faint stars is beyond the capabilities of current instruments
(see \citealt{zoutendijk20}).  As in Section~\ref{sec:tides}, a
follow-up to the N-body experiments of \citet{contenta18} may be
necessary to investigate these processes fully.

There are several caveats to the conclusions of the above studies.
First, we measured a lower luminosity for the cluster than the value
from \citet{crnojevic16} that was assumed by \citet{amorisco17} and
\citet{contenta18}.  Neither set of authors simulated clusters with a
stellar mass below $5000~\msun$, but \citet{amorisco17} found that
lighter clusters have a dynamical friction timescale longer than a
Hubble time even in a cuspy potential.  Therefore, an offset between
the cluster and the center of the galaxy does not necessarily imply
that Eri~II must be cored.  On the other hand, the smaller mass and
larger size that we determined for the cluster leave it even more
vulnerable to tidal stripping than the simulated clusters, so the
survival of the cluster in a tidal field still supports a shallow
density profile for Eri~II.  Second, the offset we determined is
between the center of the cluster and the center of the stellar
distribution of Eri~II.  The preceding discussion is contingent upon
the stars in Eri~II being centered on the dark matter halo of the
galaxy.  If the position of the cluster marks the center of the mass
distribution, and the galaxy's stars are slightly off-center, then a
cuspy profile would be allowed.  Finally, we again note that the
measured quantity from the \emph{HST} imaging is the projected
position of the cluster, which provides only a lower limit on the
three-dimensional separation between the cluster and the galaxy.  If
the geometry of the system is such that the cluster is actually at a
much larger distance and only appears to be close to the center of
Eri~II, then no conclusions about the density profile could be drawn.

If Eri~II does contain a small core, the core must have been formed
either through baryonic feedback processes or dark matter physics.
Most dark matter models that have been investigated so far do not
produce such small cores in a particularly natural way.
\citet{nishikawa20} showed that if dark matter is self-interacting,
core collapse can occur rapidly in a halo that has experienced tidal
stripping, but the orbit of Eri~II is unlikely to have brought it
close enough to the Milky Way for significant tidal mass loss.

\subsection{Implications of the Star Formation History for Dark Matter Models}
\label{sec:machos}

As initially recognized by \citet{brandt16}, the existence of a star
cluster near the center of Eri~II places limits on the composition of
dark matter in the galaxy.  If dark matter were made of massive
compact halo objects (MaCHOs) with masses on the order of tens of
solar masses, dynamical interactions between the MaCHOs and cluster
stars would heat the cluster, eventually leading to its destruction.
Lacking any quantitative SFH measurements at the time,
\citet{brandt16} considered an age range for the cluster of
$3-12$~Gyr, corresponding to upper limits of $7-2~M_{\odot}$
($14-4~M_{\odot}$) on the MaCHO mass if MaCHOs make up all (half) of
the dark matter.  This limit was tightened somewhat by
\citet{zoutendijk20} using an estimated age of 8~Gyr for the cluster.
Our results suggest that the cluster is actually $\sim13$~Gyr old,
comparable to the age of Eri~II itself, pushing the upper limit on the
mass of MaCHOs to the low end of the previously discussed range.
MaCHO dark matter fractions above 50\%\ at $M < 15~M_{\odot}$ were
also ruled out by LMC microlensing experiments
\citep[e.g.,][]{alcock01}.  More recently, microlensing of a star
crossing the giant arc in the galaxy cluster MACS~J1149 was used to
limit MaCHOs to less than 20\%\ of the dark matter in the mass range
$M < 15~M_{\odot}$ \citep{oguri18}, while constraints from lensing of
supernovae have ruled out MaCHOs as the dominant component of dark
matter from $0.01-10^{4}~M_{\odot}$ \citep{zs18}.  Because various
caveats apply to most of these constraints, limits derived from any
single technique may not be airtight, but the combination of multiple
methods appears to close the window for MaCHOs to comprise a large
fraction of dark matter in the solar masses to hundreds of solar
masses range.

The survival of the star cluster in Eri~II also constrains fuzzy dark
matter (FDM) models, where dark matter consists of an ultra-light
particle such as the axion \citep{Hu00,Hui17}.  Similar to the case
with MaCHOs, fluctuations in the FDM potential near the center of
Eri~II will act to heat the stars in the cluster over time. Assuming
an age for the cluster of 3 Gyr, \citet{marsh19} showed that FDM
particle masses below $10^{-19}$~eV are ruled out (presuming that FDM
comprises all of the dark matter).  Combined with constraints from
gravitational lensing, the Ly-$\alpha$ forest, and Milky Way satellite
galaxies, this result excludes FDM models that have been proposed to
address small-scale challenges to the $\Lambda$CDM paradigm
\citep{schutz20,Nadler20}.  Our determination of a significantly older
age for the cluster increases the lower limit on the FDM mass by
roughly a factor of 1.6 relative to the constraints derived by
\citet{marsh19}; that is, $m_{\mathrm{FDM}} \gtrsim 1.6 \times
10^{-19}$~eV.

\section{CONCLUSIONS}
\label{sec:conclusion}

We have presented an analysis of deep \emph{HST} imaging of the Milky
Way satellite galaxy Eri~II to determine its star formation history
and structure.  The data extend well below the main sequence turnoff
of the galaxy for the first time, enabling improved age measurements.
We fit the star formation history with models consisting of one and
two bursts of star formation, subject to the constraint that they
match the spectroscopic metallicity distribution from \citet{li17}.
We found that Eri~II is dominated by a very old ($\sim13$~Gyr) stellar
population containing $>80\%$ of its stars, similar to the closer and
lower-luminosity ultra-faint dwarfs that have been studied previously
\citep{brown14}.  The star formation rate in Eri~II dropped sharply by
$z \approx 6$.  It is possible that much lower rates of star formation
persisted for up to a few Gyr beyond that time, but we did not detect
such a population at a statistically significant level.  There is no
evidence for stars younger than 8~Gyr in Eri~II.  Although the orbit
of Eri~II is not strongly constrained by the available kinematic data,
comparison with simulations suggests that Eri~II is unlikely to have
approached the Milky Way early enough for environmental processes to
be responsible for its quenching.  We therefore conclude that star
formation in Eri~II was shut off by reionization.

Taking advantage of the combination of high angular resolution and
deep photometry, we also measured the structure of Eri~II, focusing
especially on its star cluster.  We determined that the cluster has a
half-light radius of $15 \pm 1$~pc, and it contains 1.8\%\ of the
total stars in Eri~II, corresponding to an absolute magnitude of
$M_{V} = -2.7 \pm 0.3$ for the Eri~II luminosity measured by
\citet{crnojevic16}.  This size is larger and the luminosity is
smaller than previously estimated.  The cluster is significantly
elongated ($e = 0.31^{+0.05}_{-0.06}$) and its position angle is
aligned with that of Eri~II to an accuracy of $3 \pm 6$~degrees.
Moreover, we showed that there is a small but significant offset of
$23 \pm 3$~pc between the center of the cluster and the center of the
galaxy.  The size, elongation, and position suggest that tidal forces
are responsible for shaping the cluster, but the origin of the
cluster's orientation is less clear.  The stellar population of the
cluster appears qualitatively consistent with that of Eri~II itself,
suggesting a likely age of $\sim13$~Gyr.

Confirmation that the cluster is not located at the center of Eri~II
provides some insight into the distribution of dark matter within the
galaxy.  Numerical simulations have shown that in a dark matter halo
with a cuspy density profile, clusters rapidly suffer tidal
destruction and (if not destroyed) are dragged to the center by
dynamical friction \citep[e.g.,][]{goerdt06,amorisco17,contenta18}.
The survival of the cluster in Eri~II and its off-center position
therefore indicate a shallow density profile.  However, the 23~pc
offset of the cluster from the center of Eri~II suggests a
surprisingly small dark matter core of size $\sim50-75$~pc.  The
implications of such a core for dark matter models remain to be
explored.

\acknowledgements{This publication is based upon work supported by
  Program number HST-GO-14234, provided by NASA through a grant from
  the Space Telescope Science Institute, which is operated by the
  Association of Universities for Research in Astronomy, Incorporated,
  under NASA contract NAS5-26555.  J.D.S. was also partially supported
  by the National Science Foundation under grant AST-1714873.
  J.S. acknowledges support from NSF grant AST-1812856 and the Packard
  Foundation.  We thank Mike Boylan-Kolchin, Xiaolong Du, Alex Ji,
  Manoj Kaplinghat, Alan McConnachie, Andrew Pace, and Ian Thompson
  for helpful conversations about various aspects of this work.  We
  also thank the anonymous referee for helpful suggestions.  Contour
  plots were generated using \code{corner.py} \citep{corner}.  This
  work has made use of data from the European Space Agency (ESA)
  mission {\it Gaia} (\url{https://www.cosmos.esa.int/gaia}),
  processed by the {\it Gaia} Data Processing and Analysis Consortium
  (DPAC,
  \url{https://www.cosmos.esa.int/web/gaia/dpac/consortium}). Funding
  for the DPAC has been provided by national institutions, in
  particular the institutions participating in the {\it Gaia}
  Multilateral Agreement.  This research has made use of NASA's
  Astrophysics Data System Bibliographic Services.}

{\it Facilities:} \facility{HST (ACS)}

{\it Software:} DAOPHOT-II \citep{stetson87}, numpy \citep{van2011numpy}, matplotlib \citep{Hunter:2007}, astropy \citep{astropy}, emcee \citep{fm13}, galpy \citep{galpy}, corner \citep{corner}

\bibliographystyle{apj}
\bibliography{main}{}

\begin{thebibliography}{}
\expandafter\ifx\csname natexlab\endcsname\relax\def\natexlab#1{#1}\fi

\bibitem[{{Adams} \& {Oosterloo}(2018)}]{adams18}
{Adams}, E. A.~K., \& {Oosterloo}, T.~A. 2018, \aap, 612, A26

\bibitem[{{Albers} {et~al.}(2019){Albers}, {Weisz}, {Cole}, {Dolphin},
  {Skillman}, {Williams}, {Boylan-Kolchin}, {Bullock}, {Dalcanton}, {Hopkins},
  {Leaman}, {McConnachie}, {Vogelsberger}, \& {Wetzel}}]{albers19}
{Albers}, S.~M., {Weisz}, D.~R., {Cole}, A.~A., {et~al.} 2019, \mnras, 490,
  5538

\bibitem[{{Alcock} {et~al.}(2001){Alcock}, {Allsman}, {Alves}, {Axelrod},
  {Becker}, {Bennett}, {Cook}, {Dalal}, {Drake}, {Freeman}, {Geha}, {Griest},
  {Lehner}, {Marshall}, {Minniti}, {Nelson}, {Peterson}, {Popowski}, {Pratt},
  {Quinn}, {Stubbs}, {Sutherland}, {Tomaney}, {Vand ehei}, \&
  {Welch}}]{alcock01}
{Alcock}, C., {Allsman}, R.~A., {Alves}, D.~R., {et~al.} 2001, \apjl, 550, L169

\bibitem[{{Amorisco}(2017)}]{amorisco17}
{Amorisco}, N.~C. 2017, \apj, 844, 64

\bibitem[{{Astropy Collaboration} {et~al.}(2018){Astropy Collaboration},
  {Price-Whelan}, {Sip{\H{o}}cz}, {G{\"u}nther}, {Lim}, {Crawford}, {Conseil},
  {Shupe}, {Craig}, {Dencheva}, {Ginsburg}, {Vand erPlas}, {Bradley},
  {P{\'e}rez-Su{\'a}rez}, {de Val-Borro}, {Aldcroft}, {Cruz}, {Robitaille},
  {Tollerud}, {Ardelean}, {Babej}, {Bach}, {Bachetti}, {Bakanov}, {Bamford},
  {Barentsen}, {Barmby}, {Baumbach}, {Berry}, {Biscani}, {Boquien}, {Bostroem},
  {Bouma}, {Brammer}, {Bray}, {Breytenbach}, {Buddelmeijer}, {Burke},
  {Calderone}, {Cano Rodr{\'\i}guez}, {Cara}, {Cardoso}, {Cheedella}, {Copin},
  {Corrales}, {Crichton}, {D'Avella}, {Deil}, {Depagne}, {Dietrich}, {Donath},
  {Droettboom}, {Earl}, {Erben}, {Fabbro}, {Ferreira}, {Finethy}, {Fox},
  {Garrison}, {Gibbons}, {Goldstein}, {Gommers}, {Greco}, {Greenfield},
  {Groener}, {Grollier}, {Hagen}, {Hirst}, {Homeier}, {Horton}, {Hosseinzadeh},
  {Hu}, {Hunkeler}, {Ivezi{\'c}}, {Jain}, {Jenness}, {Kanarek}, {Kendrew},
  {Kern}, {Kerzendorf}, {Khvalko}, {King}, {Kirkby}, {Kulkarni}, {Kumar},
  {Lee}, {Lenz}, {Littlefair}, {Ma}, {Macleod}, {Mastropietro}, {McCully},
  {Montagnac}, {Morris}, {Mueller}, {Mumford}, {Muna}, {Murphy}, {Nelson},
  {Nguyen}, {Ninan}, {N{\"o}the}, {Ogaz}, {Oh}, {Parejko}, {Parley}, {Pascual},
  {Patil}, {Patil}, {Plunkett}, {Prochaska}, {Rastogi}, {Reddy Janga},
  {Sabater}, {Sakurikar}, {Seifert}, {Sherbert}, {Sherwood-Taylor}, {Shih},
  {Sick}, {Silbiger}, {Singanamalla}, {Singer}, {Sladen}, {Sooley},
  {Sornarajah}, {Streicher}, {Teuben}, {Thomas}, {Tremblay}, {Turner},
  {Terr{\'o}n}, {van Kerkwijk}, {de la Vega}, {Watkins}, {Weaver}, {Whitmore},
  {Woillez}, {Zabalza}, \& {Astropy Contributors}}]{astropy}
{Astropy Collaboration}, {Price-Whelan}, A.~M., {Sip{\H{o}}cz}, B.~M., {et~al.}
  2018, \aj, 156, 123

\bibitem[{{Bailin} \& {Ford}(2007)}]{bf07}
{Bailin}, J., \& {Ford}, A. 2007, \mnras, 375, L41

\bibitem[{{Bailyn}(1995)}]{bailyn95}
{Bailyn}, C.~D. 1995, \araa, 33, 133

\bibitem[{{Battaglia} {et~al.}(2011){Battaglia}, {Tolstoy}, {Helmi}, {Irwin},
  {Parisi}, {Hill}, \& {Jablonka}}]{battaglia11}
{Battaglia}, G., {Tolstoy}, E., {Helmi}, A., {et~al.} 2011, \mnras, 411, 1013

\bibitem[{{Bechtol} {et~al.}(2015){Bechtol}, {Drlica-Wagner}, {Balbinot},
  {Pieres}, {Simon}, {Yanny}, {Santiago}, {Wechsler}, {Frieman}, {Walker},
  {Williams}, {Rozo}, {Rykoff}, {Queiroz}, {Luque}, {Benoit-L{\'e}vy},
  {Tucker}, {Sevilla}, {Gruendl}, {da Costa}, {Fausti Neto}, {Maia}, {Abbott},
  {Allam}, {Armstrong}, {Bauer}, {Bernstein}, {Bernstein}, {Bertin}, {Brooks},
  {Buckley-Geer}, {Burke}, {Carnero Rosell}, {Castander}, {Covarrubias},
  {D'Andrea}, {DePoy}, {Desai}, {Diehl}, {Eifler}, {Estrada}, {Evrard},
  {Fernandez}, {Finley}, {Flaugher}, {Gaztanaga}, {Gerdes}, {Girardi},
  {Gladders}, {Gruen}, {Gutierrez}, {Hao}, {Honscheid}, {Jain}, {James},
  {Kent}, {Kron}, {Kuehn}, {Kuropatkin}, {Lahav}, {Li}, {Lin}, {Makler},
  {March}, {Marshall}, {Martini}, {Merritt}, {Miller}, {Miquel}, {Mohr},
  {Neilsen}, {Nichol}, {Nord}, {Ogando}, {Peoples}, {Petravick}, {Plazas},
  {Romer}, {Roodman}, {Sako}, {Sanchez}, {Scarpine}, {Schubnell}, {Smith},
  {Soares-Santos}, {Sobreira}, {Suchyta}, {Swanson}, {Tarle}, {Thaler},
  {Thomas}, {Wester}, {Zuntz}, \& {DES Collaboration}}]{bechtol15}
{Bechtol}, K., {Drlica-Wagner}, A., {Balbinot}, E., {et~al.} 2015, \apj, 807,
  50

\bibitem[{{Belokurov} {et~al.}(2007){Belokurov}, {Zucker}, {Evans}, {Kleyna},
  {Koposov}, {Hodgkin}, {Irwin}, {Gilmore}, {Wilkinson}, {Fellhauer},
  {Bramich}, {Hewett}, {Vidrih}, {De Jong}, {Smith}, {Rix}, {Bell}, {Wyse},
  {Newberg}, {Mayeur}, {Yanny}, {Rockosi}, {Gnedin}, {Schneider}, {Beers},
  {Barentine}, {Brewington}, {Brinkmann}, {Harvanek}, {Kleinman}, {Krzesinski},
  {Long}, {Nitta}, \& {Snedden}}]{belokurov07}
{Belokurov}, V., {Zucker}, D.~B., {Evans}, N.~W., {et~al.} 2007, \apj, 654, 897

\bibitem[{{Besla} {et~al.}(2007){Besla}, {Kallivayalil}, {Hernquist},
  {Robertson}, {Cox}, {van der Marel}, \& {Alcock}}]{besla07}
{Besla}, G., {Kallivayalil}, N., {Hernquist}, L., {et~al.} 2007, \apj, 668, 949

\bibitem[{{Binney} \& {Tremaine}(2008)}]{bt08}
{Binney}, J., \& {Tremaine}, S. 2008, {Galactic Dynamics: Second Edition}
  (Princeton University Press)

\bibitem[{{Bla{\~n}a} {et~al.}(2020){Bla{\~n}a}, {Burkert}, {Fellhauer},
  {Schartmann}, \& {Alig}}]{blana20}
{Bla{\~n}a}, M., {Burkert}, A., {Fellhauer}, M., {Schartmann}, M., \& {Alig},
  C. 2020, \mnras, 497, 3601

\bibitem[{{Blitz} \& {Robishaw}(2000)}]{br00}
{Blitz}, L., \& {Robishaw}, T. 2000, \apj, 541, 675

\bibitem[{{Boldrini} {et~al.}(2019){Boldrini}, {Mohayaee}, \&
  {Silk}}]{boldrini19}
{Boldrini}, P., {Mohayaee}, R., \& {Silk}, J. 2019, \mnras, 485, 2546

\bibitem[{{Bovill} \& {Ricotti}(2011)}]{br11a}
{Bovill}, M.~S., \& {Ricotti}, M. 2011, \apj, 741, 17

\bibitem[{{Bovy}(2015)}]{galpy}
{Bovy}, J. 2015, \apjs, 216, 29

\bibitem[{{Brandt}(2016)}]{brandt16}
{Brandt}, T.~D. 2016, \apjl, 824, L31

\bibitem[{{Brown} {et~al.}(2005){Brown}, {Ferguson}, {Smith}, {Guhathakurta},
  {Kimble}, {Sweigart}, {Renzini}, {Rich}, \& {Vand enBerg}}]{brown05}
{Brown}, T.~M., {Ferguson}, H.~C., {Smith}, E., {et~al.} 2005, \aj, 130, 1693

\bibitem[{{Brown} {et~al.}(2012){Brown}, {Tumlinson}, {Geha}, {Kirby},
  {VandenBerg}, {Mu{\~n}oz}, {Kalirai}, {Simon}, {Avila}, {Guhathakurta},
  {Renzini}, \& {Ferguson}}]{brown12}
{Brown}, T.~M., {Tumlinson}, J., {Geha}, M., {et~al.} 2012, \apjl, 753, L21

\bibitem[{{Brown} {et~al.}(2014){Brown}, {Tumlinson}, {Geha}, {Simon},
  {Vargas}, {VandenBerg}, {Kirby}, {Kalirai}, {Avila}, {Gennaro}, {Ferguson},
  {Mu{\~n}oz}, {Guhathakurta}, \& {Renzini}}]{brown14}
---. 2014, \apj, 796, 91

\bibitem[{{Carlin} \& {Sand}(2018)}]{carlin18}
{Carlin}, J.~L., \& {Sand}, D.~J. 2018, \apj, 865, 7

\bibitem[{{Ciotti} \& {Bertin}(1999)}]{cb99}
{Ciotti}, L., \& {Bertin}, G. 1999, \aap, 352, 447

\bibitem[{{Clementini} {et~al.}(2012){Clementini}, {Cignoni}, {Contreras
  Ramos}, {Federici}, {Ripepi}, {Marconi}, {Tosi}, \& {Musella}}]{clementini12}
{Clementini}, G., {Cignoni}, M., {Contreras Ramos}, R., {et~al.} 2012, \apj,
  756, 108

\bibitem[{{Contenta} {et~al.}(2018){Contenta}, {Balbinot}, {Petts}, {Read},
  {Gieles}, {Collins}, {Pe{\~n}arrubia}, {Delorme}, \&
  {Gualandris}}]{contenta18}
{Contenta}, F., {Balbinot}, E., {Petts}, J.~A., {et~al.} 2018, \mnras, 476,
  3124

\bibitem[{{Crnojevi{\'c}} {et~al.}(2016){Crnojevi{\'c}}, {Sand}, {Zaritsky},
  {Spekkens}, {Willman}, \& {Hargis}}]{crnojevic16}
{Crnojevi{\'c}}, D., {Sand}, D.~J., {Zaritsky}, D., {et~al.} 2016, \apjl, 824,
  L14

\bibitem[{{Cusano} {et~al.}(2016){Cusano}, {Garofalo}, {Clementini}, {Cignoni},
  {Federici}, {Marconi}, {Ripepi}, {Musella}, {Testa}, {Carini}, \&
  {Faccini}}]{cusano16}
{Cusano}, F., {Garofalo}, A., {Clementini}, G., {et~al.} 2016, \apj, 829, 26

\bibitem[{{de Blok} {et~al.}(2001){de Blok}, {McGaugh}, {Bosma}, \&
  {Rubin}}]{deblok01}
{de Blok}, W.~J.~G., {McGaugh}, S.~S., {Bosma}, A., \& {Rubin}, V.~C. 2001,
  \apjl, 552, L23

\bibitem[{{de Jong} {et~al.}(2008){de Jong}, {Harris}, {Coleman}, {Martin},
  {Bell}, {Rix}, {Hill}, {Skillman}, {Sand}, {Olszewski}, {Zaritsky},
  {Thompson}, {Giallongo}, {Ragazzoni}, {DiPaola}, {Farinato}, {Testa}, \&
  {Bechtold}}]{dejong08}
{de Jong}, J.~T.~A., {Harris}, J., {Coleman}, M.~G., {et~al.} 2008, \apj, 680,
  1112

\bibitem[{{Del Principe} {et~al.}(2005){Del Principe}, {Piersimoni}, {Bono},
  {Di Paola}, {Dolci}, \& {Marconi}}]{delprincipe05}
{Del Principe}, M., {Piersimoni}, A.~M., {Bono}, G., {et~al.} 2005, \aj, 129,
  2714

\bibitem[{{DES Collaboration} {et~al.}(2018){DES Collaboration}, {Abbott},
  {Abdalla}, {Allam}, {Amara}, {Annis}, {Asorey}, {Avila}, {Ballester},
  {Banerji}, {Barkhouse}, {Baruah}, {Baumer}, {Bechtol}, {Becker},
  {Benoit-L{\'e}vy}, {Bernstein}, {Bertin}, {Blazek}, {Bocquet}, {Brooks},
  {Brout}, {Buckley-Geer}, {Burke}, {Busti}, {Campisano}, {Cardiel-Sas},
  {Carnero Rosell}, {Carrasco Kind}, {Carretero}, {Castander}, {Cawthon},
  {Chang}, {Chen}, {Conselice}, {Costa}, {Crocce}, {Cunha}, {D'Andrea}, {da
  Costa}, {Das}, {Daues}, {Davis}, {Davis}, {De Vicente}, {DePoy}, {DeRose},
  {Desai}, {Diehl}, {Dietrich}, {Dodelson}, {Doel}, {Drlica-Wagner}, {Eifler},
  {Elliott}, {Evrard}, {Farahi}, {Fausti Neto}, {Fernandez}, {Finley},
  {Flaugher}, {Foley}, {Fosalba}, {Friedel}, {Frieman}, {Garc{\'\i}a-Bellido},
  {Gaztanaga}, {Gerdes}, {Giannantonio}, {Gill}, {Glazebrook}, {Goldstein},
  {Gower}, {Gruen}, {Gruendl}, {Gschwend}, {Gupta}, {Gutierrez}, {Hamilton},
  {Hartley}, {Hinton}, {Hislop}, {Hollowood}, {Honscheid}, {Hoyle}, {Huterer},
  {Jain}, {James}, {Jeltema}, {Johnson}, {Johnson}, {Kacprzak}, {Kent},
  {Khullar}, {Klein}, {Kovacs}, {Koziol}, {Krause}, {Kremin}, {Kron}, {Kuehn},
  {Kuhlmann}, {Kuropatkin}, {Lahav}, {Lasker}, {Li}, {Li}, {Liddle}, {Lima},
  {Lin}, {L{\'o}pez-Reyes}, {MacCrann}, {Maia}, {Maloney}, {Manera}, {March},
  {Marriner}, {Marshall}, {Martini}, {McClintock}, {McKay}, {McMahon},
  {Melchior}, {Menanteau}, {Miller}, {Miquel}, {Mohr}, {Morganson}, {Mould},
  {Neilsen}, {Nichol}, {Nogueira}, {Nord}, {Nugent}, {Nunes}, {Ogand o}, {Old},
  {Pace}, {Palmese}, {Paz-Chinch{\'o}n}, {Peiris}, {Percival}, {Petravick},
  {Plazas}, {Poh}, {Pond}, {Porredon}, {Pujol}, {Refregier}, {Reil}, {Ricker},
  {Rollins}, {Romer}, {Roodman}, {Rooney}, {Ross}, {Rykoff}, {Sako}, {Sanchez},
  {Sanchez}, {Santiago}, {Saro}, {Scarpine}, {Scolnic}, {Serrano},
  {Sevilla-Noarbe}, {Sheldon}, {Shipp}, {Silveira}, {Smith}, {Smith}, {Smith},
  {Soares-Santos}, {Sobreira}, {Song}, {Stebbins}, {Suchyta}, {Sullivan},
  {Swanson}, {Tarle}, {Thaler}, {Thomas}, {Thomas}, {Troxel}, {Tucker},
  {Vikram}, {Vivas}, {Walker}, {Wechsler}, {Weller}, {Wester}, {Wolf}, {Wu},
  {Yanny}, {Zenteno}, {Zhang}, {Zuntz}, {DES Collaboration}, {Juneau},
  {Fitzpatrick}, {Nikutta}, {Nidever}, {Olsen}, {Scott}, \& {NOAO Data
  Lab}}]{desdr1}
{DES Collaboration}, {Abbott}, T.~M.~C., {Abdalla}, F.~B., {et~al.} 2018,
  \apjs, 239, 18

\bibitem[{{Dolphin}(2002)}]{dolphin02}
{Dolphin}, A.~E. 2002, \mnras, 332, 91

\bibitem[{{Dolphin}(2012)}]{dolphin12}
---. 2012, \apj, 751, 60

\bibitem[{{Drlica-Wagner} {et~al.}(2015){Drlica-Wagner}, {Bechtol}, {Rykoff},
  {Luque}, {Queiroz}, {Mao}, {Wechsler}, {Simon}, {Santiago}, {Yanny},
  {Balbinot}, {Dodelson}, {Fausti Neto}, {James}, {Li}, {Maia}, {Marshall},
  {Pieres}, {Stringer}, {Walker}, {Abbott}, {Abdalla}, {Allam},
  {Benoit-L{\'e}vy}, {Bernstein}, {Bertin}, {Brooks}, {Buckley-Geer}, {Burke},
  {Carnero Rosell}, {Carrasco Kind}, {Carretero}, {Crocce}, {da Costa},
  {Desai}, {Diehl}, {Dietrich}, {Doel}, {Eifler}, {Evrard}, {Finley},
  {Flaugher}, {Fosalba}, {Frieman}, {Gaztanaga}, {Gerdes}, {Gruen}, {Gruendl},
  {Gutierrez}, {Honscheid}, {Kuehn}, {Kuropatkin}, {Lahav}, {Martini},
  {Miquel}, {Nord}, {Ogando}, {Plazas}, {Reil}, {Roodman}, {Sako}, {Sanchez},
  {Scarpine}, {Schubnell}, {Sevilla-Noarbe}, {Smith}, {Soares-Santos},
  {Sobreira}, {Suchyta}, {Swanson}, {Tarle}, {Tucker}, {Vikram}, {Wester},
  {Zhang}, {Zuntz}, \& {DES Collaboration}}]{drlica15}
{Drlica-Wagner}, A., {Bechtol}, K., {Rykoff}, E.~S., {et~al.} 2015, \apj, 813,
  109

\bibitem[{{Drlica-Wagner} {et~al.}(2020){Drlica-Wagner}, {Bechtol}, {Mau},
  {McNanna}, {Nadler}, {Pace}, {Li}, {Pieres}, {Rozo}, {Simon}, {Walker},
  {Wechsler}, {Abbott}, {Allam}, {Annis}, {Bertin}, {Brooks}, {Burke},
  {Rosell}, {Carrasco Kind}, {Carretero}, {Costanzi}, {da Costa}, {De Vicente},
  {Desai}, {Diehl}, {Doel}, {Eifler}, {Everett}, {Flaugher}, {Frieman},
  {Garc{\'\i}a-Bellido}, {Gaztanaga}, {Gruen}, {Gruendl}, {Gschwend},
  {Gutierrez}, {Honscheid}, {James}, {Krause}, {Kuehn}, {Kuropatkin}, {Lahav},
  {Maia}, {Marshall}, {Melchior}, {Menanteau}, {Miquel}, {Palmese}, {Plazas},
  {Sanchez}, {Scarpine}, {Schubnell}, {Serrano}, {Sevilla-Noarbe}, {Smith},
  {Suchyta}, {Tarle}, \& {DES Collaboration}}]{Drlica20}
{Drlica-Wagner}, A., {Bechtol}, K., {Mau}, S., {et~al.} 2020, \apj, 893, 47

\bibitem[{{Einasto} {et~al.}(1974){Einasto}, {Saar}, {Kaasik}, \&
  {Chernin}}]{einasto74}
{Einasto}, J., {Saar}, E., {Kaasik}, A., \& {Chernin}, A.~D. 1974, \nat, 252,
  111

\bibitem[{{Fillingham} {et~al.}(2018){Fillingham}, {Cooper}, {Boylan-Kolchin},
  {Bullock}, {Garrison-Kimmel}, \& {Wheeler}}]{fillingham18}
{Fillingham}, S.~P., {Cooper}, M.~C., {Boylan-Kolchin}, M., {et~al.} 2018,
  \mnras, 477, 4491

\bibitem[{{Flores} \& {Primack}(1994)}]{fp94}
{Flores}, R.~A., \& {Primack}, J.~R. 1994, \apjl, 427, L1

\bibitem[{{Ford} {et~al.}(2003){Ford}, {Clampin}, {Hartig}, {Illingworth},
  {Sirianni}, {Martel}, {Meurer}, {McCann}, {Sullivan}, {Bartko}, {Benitez},
  {Blakeslee}, {Bouwens}, {Broadhurst}, {Brown}, {Burrows}, {Campbell},
  {Cheng}, {Feldman}, {Franx}, {Golimowski}, {Gronwall}, {Kimble}, {Krist},
  {Lesser}, {Magee}, {Miley}, {Postman}, {Rafal}, {Rosati}, {Sparks}, {Tran},
  {Tsvetanov}, {Volmer}, {White}, \& {Woodruff}}]{acs}
{Ford}, H.~C., {Clampin}, M., {Hartig}, G.~F., {et~al.} 2003, Society of
  Photo-Optical Instrumentation Engineers (SPIE) Conference Series, Vol. 4854,
  {Overview of the Advanced Camera for Surveys on-orbit performance}, ed. J.~C.
  {Blades} \& O.~H.~W. {Siegmund}, 81--94

\bibitem[{Foreman-Mackey(2016)}]{corner}
Foreman-Mackey, D. 2016, The Journal of Open Source Software, 24,
  doi:10.21105/joss.00024

\bibitem[{{Foreman-Mackey} {et~al.}(2013){Foreman-Mackey}, {Hogg}, {Lang}, \&
  {Goodman}}]{fm13}
{Foreman-Mackey}, D., {Hogg}, D.~W., {Lang}, D., \& {Goodman}, J. 2013, \pasp,
  125, 306

\bibitem[{{Fritz} {et~al.}(2018){Fritz}, {Battaglia}, {Pawlowski},
  {Kallivayalil}, {van der Marel}, {Sohn}, {Brook}, \& {Besla}}]{fritz18}
{Fritz}, T.~K., {Battaglia}, G., {Pawlowski}, M.~S., {et~al.} 2018, \aap, 619,
  A103

\bibitem[{{Gaia Collaboration} {et~al.}(2016){Gaia Collaboration}, {Prusti},
  {de Bruijne}, {Brown}, {Vallenari}, {Babusiaux}, {Bailer-Jones}, {Bastian},
  {Biermann}, {Evans}, \& et~al.}]{gaia16a}
{Gaia Collaboration}, {Prusti}, T., {de Bruijne}, J.~H.~J., {et~al.} 2016,
  \aap, 595, A1

\bibitem[{{Gaia Collaboration} {et~al.}(2018){Gaia Collaboration}, {Brown},
  {Vallenari}, {Prusti}, {de Bruijne}, {Babusiaux}, {Bailer-Jones}, {Biermann},
  {Evans}, {Eyer}, {Jansen}, {Jordi}, {Klioner}, {Lammers}, {Lindegren},
  {Luri}, {Mignard}, {Panem}, {Pourbaix}, {Randich}, {Sartoretti}, {Siddiqui},
  {Soubiran}, {van Leeuwen}, {Walton}, {Arenou}, {Bastian}, {Cropper},
  {Drimmel}, {Katz}, {Lattanzi}, {Bakker}, {Cacciari}, {Casta{\~n}eda},
  {Chaoul}, {Cheek}, {De Angeli}, {Fabricius}, {Guerra}, {Holl}, {Masana},
  {Messineo}, {Mowlavi}, {Nienartowicz}, {Panuzzo}, {Portell}, {Riello},
  {Seabroke}, {Tanga}, {Th{\'e}venin}, {Gracia-Abril}, {Comoretto},
  {Garcia-Reinaldos}, {Teyssier}, {Altmann}, {Andrae}, {Audard},
  {Bellas-Velidis}, {Benson}, {Berthier}, {Blomme}, {Burgess}, {Busso},
  {Carry}, {Cellino}, {Clementini}, {Clotet}, {Creevey}, {Davidson}, {De
  Ridder}, {Delchambre}, {Dell'Oro}, {Ducourant},
  {Fern{\'a}ndez-Hern{\'a}ndez}, {Fouesneau}, {Fr{\'e}mat}, {Galluccio},
  {Garc{\'\i}a-Torres}, {Gonz{\'a}lez-N{\'u}{\~n}ez}, {Gonz{\'a}lez-Vidal},
  {Gosset}, {Guy}, {Halbwachs}, {Hambly}, {Harrison}, {Hern{\'a}ndez},
  {Hestroffer}, {Hodgkin}, {Hutton}, {Jasniewicz}, {Jean-Antoine-Piccolo},
  {Jordan}, {Korn}, {Krone-Martins}, {Lanzafame}, {Lebzelter}, {L{\"o}ffler},
  {Manteiga}, {Marrese}, {Mart{\'\i}n-Fleitas}, {Moitinho}, {Mora}, {Muinonen},
  {Osinde}, {Pancino}, {Pauwels}, {Petit}, {Recio-Blanco}, {Richards},
  {Rimoldini}, {Robin}, {Sarro}, {Siopis}, {Smith}, {Sozzetti}, {S{\"u}veges},
  {Torra}, {van Reeven}, {Abbas}, {Abreu Aramburu}, {Accart}, {Aerts},
  {Altavilla}, {{\'A}lvarez}, {Alvarez}, {Alves}, {Anderson}, {Andrei},
  {Anglada Varela}, {Antiche}, {Antoja}, {Arcay}, {Astraatmadja}, {Bach},
  {Baker}, {Balaguer-N{\'u}{\~n}ez}, {Balm}, {Barache}, {Barata}, {Barbato},
  {Barblan}, {Barklem}, {Barrado}, {Barros}, {Barstow}, {Bartholom{\'e}
  Mu{\~n}oz}, {Bassilana}, {Becciani}, {Bellazzini}, {Berihuete}, {Bertone},
  {Bianchi}, {Bienaym{\'e}}, {Blanco-Cuaresma}, {Boch}, {Boeche}, {Bombrun},
  {Borrachero}, {Bossini}, {Bouquillon}, {Bourda}, {Bragaglia}, {Bramante},
  {Breddels}, {Bressan}, {Brouillet}, {Br{\"u}semeister}, {Brugaletta},
  {Bucciarelli}, {Burlacu}, {Busonero}, {Butkevich}, {Buzzi}, {Caffau},
  {Cancelliere}, {Cannizzaro}, {Cantat-Gaudin}, {Carballo}, {Carlucci},
  {Carrasco}, {Casamiquela}, {Castellani}, {Castro-Ginard}, {Charlot},
  {Chemin}, {Chiavassa}, {Cocozza}, {Costigan}, {Cowell}, {Crifo}, {Crosta},
  {Crowley}, {Cuypers}, {Dafonte}, {Damerdji}, {Dapergolas}, {David}, {David},
  {de Laverny}, {De Luise}, {De March}, {de Martino}, {de Souza}, {de Torres},
  {Debosscher}, {del Pozo}, {Delbo}, {Delgado}, {Delgado}, {Di Matteo},
  {Diakite}, {Diener}, {Distefano}, {Dolding}, {Drazinos}, {Dur{\'a}n},
  {Edvardsson}, {Enke}, {Eriksson}, {Esquej}, {Eynard Bontemps}, {Fabre},
  {Fabrizio}, {Faigler}, {Falc{\~a}o}, {Farr{\`a}s Casas}, {Federici},
  {Fedorets}, {Fernique}, {Figueras}, {Filippi}, {Findeisen}, {Fonti},
  {Fraile}, {Fraser}, {Fr{\'e}zouls}, {Gai}, {Galleti}, {Garabato},
  {Garc{\'\i}a-Sedano}, {Garofalo}, {Garralda}, {Gavel}, {Gavras}, {Gerssen},
  {Geyer}, {Giacobbe}, {Gilmore}, {Girona}, {Giuffrida}, {Glass}, {Gomes},
  {Granvik}, {Gueguen}, {Guerrier}, {Guiraud}, {Guti{\'e}rrez-S{\'a}nchez},
  {Haigron}, {Hatzidimitriou}, {Hauser}, {Haywood}, {Heiter}, {Helmi}, {Heu},
  {Hilger}, {Hobbs}, {Hofmann}, {Holland}, {Huckle}, {Hypki}, {Icardi},
  {Jan{\ss}en}, {Jevardat de Fombelle}, {Jonker}, {Juh{\'a}sz}, {Julbe},
  {Karampelas}, {Kewley}, {Klar}, {Kochoska}, {Kohley}, {Kolenberg},
  {Kontizas}, {Kontizas}, {Koposov}, {Kordopatis}, {Kostrzewa-Rutkowska},
  {Koubsky}, {Lambert}, {Lanza}, {Lasne}, {Lavigne}, {Le Fustec}, {Le
  Poncin-Lafitte}, {Lebreton}, {Leccia}, {Leclerc}, {Lecoeur-Taibi},
  {Lenhardt}, {Leroux}, {Liao}, {Licata}, {Lindstr{\o}m}, {Lister}, {Livanou},
  {Lobel}, {L{\'o}pez}, {Managau}, {Mann}, {Mantelet}, {Marchal}, {Marchant},
  {Marconi}, {Marinoni}, {Marschalk{\'o}}, {Marshall}, {Martino}, {Marton},
  {Mary}, {Massari}, {Matijevi{\v{c}}}, {Mazeh}, {McMillan}, {Messina},
  {Michalik}, {Millar}, {Molina}, {Molinaro}, {Moln{\'a}r}, {Montegriffo},
  {Mor}, {Morbidelli}, {Morel}, {Morris}, {Mulone}, {Muraveva}, {Musella},
  {Nelemans}, {Nicastro}, {Noval}, {O'Mullane}, {Ord{\'e}novic},
  {Ord{\'o}{\~n}ez-Blanco}, {Osborne}, {Pagani}, {Pagano}, {Pailler},
  {Palacin}, {Palaversa}, {Panahi}, {Pawlak}, {Piersimoni}, {Pineau}, {Plachy},
  {Plum}, {Poggio}, {Poujoulet}, {Pr{\v{s}}a}, {Pulone}, {Racero}, {Ragaini},
  {Rambaux}, {Ramos-Lerate}, {Regibo}, {Reyl{\'e}}, {Riclet}, {Ripepi}, {Riva},
  {Rivard}, {Rixon}, {Roegiers}, {Roelens}, {Romero-G{\'o}mez}, {Rowell},
  {Royer}, {Ruiz-Dern}, {Sadowski}, {Sagrist{\`a} Sell{\'e}s}, {Sahlmann},
  {Salgado}, {Salguero}, {Sanna}, {Santana-Ros}, {Sarasso}, {Savietto},
  {Schultheis}, {Sciacca}, {Segol}, {Segovia}, {S{\'e}gransan}, {Shih},
  {Siltala}, {Silva}, {Smart}, {Smith}, {Solano}, {Solitro}, {Sordo}, {Soria
  Nieto}, {Souchay}, {Spagna}, {Spoto}, {Stampa}, {Steele},
  {Steidelm{\"u}ller}, {Stephenson}, {Stoev}, {Suess}, {Surdej}, {Szabados},
  {Szegedi-Elek}, {Tapiador}, {Taris}, {Tauran}, {Taylor}, {Teixeira},
  {Terrett}, {Teyssand ier}, {Thuillot}, {Titarenko}, {Torra Clotet}, {Turon},
  {Ulla}, {Utrilla}, {Uzzi}, {Vaillant}, {Valentini}, {Valette}, {van Elteren},
  {Van Hemelryck}, {van Leeuwen}, {Vaschetto}, {Vecchiato}, {Veljanoski},
  {Viala}, {Vicente}, {Vogt}, {von Essen}, {Voss}, {Votruba}, {Voutsinas},
  {Walmsley}, {Weiler}, {Wertz}, {Wevers}, {Wyrzykowski}, {Yoldas},
  {{\v{Z}}erjal}, {Ziaeepour}, {Zorec}, {Zschocke}, {Zucker}, {Zurbach}, \&
  {Zwitter}}]{gaiadr2brown}
{Gaia Collaboration}, {Brown}, A.~G.~A., {Vallenari}, A., {et~al.} 2018, \aap,
  616, A1

\bibitem[{{Gallart} {et~al.}(1999){Gallart}, {Freedman}, {Aparicio},
  {Bertelli}, \& {Chiosi}}]{gallart99}
{Gallart}, C., {Freedman}, W.~L., {Aparicio}, A., {Bertelli}, G., \& {Chiosi},
  C. 1999, \aj, 118, 2245

\bibitem[{{Garrison-Kimmel} {et~al.}(2014){Garrison-Kimmel}, {Boylan-Kolchin},
  {Bullock}, \& {Lee}}]{gk14}
{Garrison-Kimmel}, S., {Boylan-Kolchin}, M., {Bullock}, J.~S., \& {Lee}, K.
  2014, \mnras, 438, 2578

\bibitem[{{Geha} {et~al.}(2015){Geha}, {Weisz}, {Grocholski}, {Dolphin}, {van
  der Marel}, \& {Guhathakurta}}]{geha15}
{Geha}, M., {Weisz}, D., {Grocholski}, A., {et~al.} 2015, \apj, 811, 114

\bibitem[{{Geha} {et~al.}(2013){Geha}, {Brown}, {Tumlinson}, {Kalirai},
  {Simon}, {Kirby}, {Vand enBerg}, {Mu{\~n}oz}, {Avila}, {Guhathakurta}, \&
  {Ferguson}}]{geha13}
{Geha}, M., {Brown}, T.~M., {Tumlinson}, J., {et~al.} 2013, \apj, 771, 29

\bibitem[{{Geha} {et~al.}(2017){Geha}, {Wechsler}, {Mao}, {Tollerud}, {Weiner},
  {Bernstein}, {Hoyle}, {Marchi}, {Marshall}, {Mu{\~n}oz}, \& {Lu}}]{geha17}
{Geha}, M., {Wechsler}, R.~H., {Mao}, Y.-Y., {et~al.} 2017, \apj, 847, 4

\bibitem[{{Goerdt} {et~al.}(2006){Goerdt}, {Moore}, {Read}, {Stadel}, \&
  {Zemp}}]{goerdt06}
{Goerdt}, T., {Moore}, B., {Read}, J.~I., {Stadel}, J., \& {Zemp}, M. 2006,
  \mnras, 368, 1073

\bibitem[{{Grcevich} \& {Putman}(2009)}]{gp09}
{Grcevich}, J., \& {Putman}, M.~E. 2009, \apj, 696, 385

\bibitem[{{Harris}(1996)}]{Harris96}
{Harris}, W.~E. 1996, \aj, 112, 1487

\bibitem[{{Hesser} {et~al.}(1984){Hesser}, {McClure}, {Hawarden}, {Cannon},
  {von Rudloff}, {Kruger}, \& {Egles}}]{hesser84}
{Hesser}, J.~E., {McClure}, R.~D., {Hawarden}, T.~G., {et~al.} 1984, \pasp, 96,
  406

\bibitem[{{Hidalgo} {et~al.}(2011){Hidalgo}, {Aparicio}, {Skillman}, {Monelli},
  {Gallart}, {Cole}, {Dolphin}, {Weisz}, {Bernard}, {Cassisi}, {Mayer},
  {Stetson}, {Tolstoy}, \& {Ferguson}}]{hidalgo11}
{Hidalgo}, S.~L., {Aparicio}, A., {Skillman}, E., {et~al.} 2011, \apj, 730, 14

\bibitem[{{Hu} {et~al.}(2000){Hu}, {Barkana}, \& {Gruzinov}}]{Hu00}
{Hu}, W., {Barkana}, R., \& {Gruzinov}, A. 2000, \prl, 85, 1158

\bibitem[{{Hui} {et~al.}(2017){Hui}, {Ostriker}, {Tremaine}, \&
  {Witten}}]{Hui17}
{Hui}, L., {Ostriker}, J.~P., {Tremaine}, S., \& {Witten}, E. 2017, \prd, 95,
  043541

\bibitem[{Hunter(2007)}]{Hunter:2007}
Hunter, J.~D. 2007, Computing In Science \& Engineering, 9, 90

\bibitem[{Husser(2012)}]{husser12}
Husser, T.-O. 2012, Ph.D. Thesis, Universit{\"a}tsverlag G{\"o}ttingen,
  doi:10.17875/gup2012-87

\bibitem[{{Irwin} {et~al.}(2007){Irwin}, {Belokurov}, {Evans}, {Ryan-Weber},
  {de Jong}, {Koposov}, {Zucker}, {Hodgkin}, {Gilmore}, {Prema}, {Hebb},
  {Begum}, {Fellhauer}, {Hewett}, {Kennicutt}, {Wilkinson}, {Bramich},
  {Vidrih}, {Rix}, {Beers}, {Barentine}, {Brewington}, {Harvanek},
  {Krzesinski}, {Long}, {Nitta}, \& {Snedden}}]{irwin07}
{Irwin}, M.~J., {Belokurov}, V., {Evans}, N.~W., {et~al.} 2007, \apjl, 656, L13

\bibitem[{{Kallivayalil} {et~al.}(2013){Kallivayalil}, {van der Marel},
  {Besla}, {Anderson}, \& {Alcock}}]{kallivayalil13}
{Kallivayalil}, N., {van der Marel}, R.~P., {Besla}, G., {Anderson}, J., \&
  {Alcock}, C. 2013, \apj, 764, 161

\bibitem[{{Kaluzny} {et~al.}(2013){Kaluzny}, {Rozyczka}, {Pych}, {Krzeminski},
  {Zloczewski}, {Narloch}, \& {Thompson}}]{kaluzny13}
{Kaluzny}, J., {Rozyczka}, M., {Pych}, W., {et~al.} 2013, \actaa, 63, 309

\bibitem[{{Kaur} \& {Sridhar}(2018)}]{kaur18}
{Kaur}, K., \& {Sridhar}, S. 2018, \apj, 868, 134

\bibitem[{{King}(1962)}]{king62}
{King}, I. 1962, \aj, 67, 471

\bibitem[{{Koposov} {et~al.}(2015){Koposov}, {Belokurov}, {Torrealba}, \&
  {Evans}}]{koposov15}
{Koposov}, S.~E., {Belokurov}, V., {Torrealba}, G., \& {Evans}, N.~W. 2015,
  \apj, 805, 130

\bibitem[{{Kuzio de Naray} {et~al.}(2008){Kuzio de Naray}, {McGaugh}, \& {de
  Blok}}]{kuzio08}
{Kuzio de Naray}, R., {McGaugh}, S.~S., \& {de Blok}, W.~J.~G. 2008, \apj, 676,
  920

\bibitem[{{Leiner} {et~al.}(2016){Leiner}, {Mathieu}, {Stello}, {Vand erburg},
  \& {Sandquist}}]{leiner16}
{Leiner}, E., {Mathieu}, R.~D., {Stello}, D., {Vand erburg}, A., \&
  {Sandquist}, E. 2016, \apjl, 832, L13

\bibitem[{{Leung} {et~al.}(2020){Leung}, {Leaman}, {van de Ven}, \&
  {Battaglia}}]{leung20}
{Leung}, G. Y.~C., {Leaman}, R., {van de Ven}, G., \& {Battaglia}, G. 2020,
  \mnras, 493, 320

\bibitem[{{Li} {et~al.}(2017){Li}, {Simon}, {Drlica-Wagner}, {Bechtol}, {Wang},
  {Garc{\'{\i}}a-Bellido}, {Frieman}, {Marshall}, {James}, {Strigari}, {Pace},
  {Balbinot}, {Zhang}, {Abbott}, {Allam}, {Benoit-L{\'e}vy}, {Bernstein},
  {Bertin}, {Brooks}, {Burke}, {Carnero Rosell}, {Carrasco Kind}, {Carretero},
  {Cunha}, {D'Andrea}, {da Costa}, {DePoy}, {Desai}, {Diehl}, {Eifler},
  {Flaugher}, {Goldstein}, {Gruen}, {Gruendl}, {Gschwend}, {Gutierrez},
  {Krause}, {Kuehn}, {Lin}, {Maia}, {March}, {Menanteau}, {Miquel}, {Plazas},
  {Romer}, {Sanchez}, {Santiago}, {Schubnell}, {Sevilla-Noarbe}, {Smith},
  {Sobreira}, {Suchyta}, {Tarle}, {Thomas}, {Tucker}, {Walker}, {Wechsler},
  {Wester}, {Yanny}, \& {(DES Collaboration}}]{li17}
{Li}, T.~S., {Simon}, J.~D., {Drlica-Wagner}, A., {et~al.} 2017, \apj, 838, 8

\bibitem[{{Lin} \& {Faber}(1983)}]{lf83}
{Lin}, D.~N.~C., \& {Faber}, S.~M. 1983, \apjl, 266, L21

\bibitem[{{Lindegren} {et~al.}(2018){Lindegren}, {Hern{\'a}ndez}, {Bombrun},
  {Klioner}, {Bastian}, {Ramos-Lerate}, {de Torres}, {Steidelm{\"u}ller},
  {Stephenson}, {Hobbs}, {Lammers}, {Biermann}, {Geyer}, {Hilger}, {Michalik},
  {Stampa}, {McMillan}, {Casta{\~n}eda}, {Clotet}, {Comoretto}, {Davidson},
  {Fabricius}, {Gracia}, {Hambly}, {Hutton}, {Mora}, {Portell}, {van Leeuwen},
  {Abbas}, {Abreu}, {Altmann}, {Andrei}, {Anglada}, {Balaguer-N{\'u}{\~n}ez},
  {Barache}, {Becciani}, {Bertone}, {Bianchi}, {Bouquillon}, {Bourda},
  {Br{\"u}semeister}, {Bucciarelli}, {Busonero}, {Buzzi}, {Cancelliere},
  {Carlucci}, {Charlot}, {Cheek}, {Crosta}, {Crowley}, {de Bruijne}, {de
  Felice}, {Drimmel}, {Esquej}, {Fienga}, {Fraile}, {Gai}, {Garralda},
  {Gonz{\'a}lez-Vidal}, {Guerra}, {Hauser}, {Hofmann}, {Holl}, {Jordan},
  {Lattanzi}, {Lenhardt}, {Liao}, {Licata}, {Lister}, {L{\"o}ffler},
  {Marchant}, {Martin-Fleitas}, {Messineo}, {Mignard}, {Morbidelli}, {Poggio},
  {Riva}, {Rowell}, {Salguero}, {Sarasso}, {Sciacca}, {Siddiqui}, {Smart},
  {Spagna}, {Steele}, {Taris}, {Torra}, {van Elteren}, {van Reeven}, \&
  {Vecchiato}}]{gaiadr2lindegren}
{Lindegren}, L., {Hern{\'a}ndez}, J., {Bombrun}, A., {et~al.} 2018, \aap, 616,
  A2

\bibitem[{{Marsh} \& {Niemeyer}(2019)}]{marsh19}
{Marsh}, D. J.~E., \& {Niemeyer}, J.~C. 2019, \prl, 123, 051103

\bibitem[{{Martin} {et~al.}(2008){Martin}, {de Jong}, \& {Rix}}]{martin08}
{Martin}, N.~F., {de Jong}, J.~T.~A., \& {Rix}, H.-W. 2008, \apj, 684, 1075

\bibitem[{{Martin} {et~al.}(2006){Martin}, {Ibata}, {Irwin}, {Chapman},
  {Lewis}, {Ferguson}, {Tanvir}, \& {McConnachie}}]{martin06}
{Martin}, N.~F., {Ibata}, R.~A., {Irwin}, M.~J., {et~al.} 2006, \mnras, 371,
  1983

\bibitem[{{Mart{\'\i}nez-V{\'a}zquez}
  {et~al.}(2019){Mart{\'\i}nez-V{\'a}zquez}, {Vivas}, {Gurevich}, {Walker},
  {McCarthy}, {Pace}, {Stringer}, {Santiago}, {Hounsell}, {Macri}, {Li},
  {Bechtol}, {Riley}, {Kim}, {Simon}, {Drlica-Wagner}, {Nadler}, {Marshall},
  {Annis}, {Avila}, {Bertin}, {Brooks}, {Buckley-Geer}, {Burke}, {Carnero
  Rosell}, {Carrasco Kind}, {da Costa}, {De Vicente}, {Desai}, {Diehl}, {Doel},
  {Everett}, {Frieman}, {Garc{\'\i}a-Bellido}, {Gaztanaga}, {Gruen}, {Gruendl},
  {Gschwend}, {Gutierrez}, {Hollowood}, {Honscheid}, {James}, {Kuehn},
  {Kuropatkin}, {Maia}, {Menanteau}, {Miller}, {Miquel}, {Paz-Chinch{\'o}n},
  {Plazas}, {Sanchez}, {Scarpine}, {Serrano}, {Sevilla-Noarbe}, {Smith},
  {Soares-Santos}, {Sobreira}, {Swanson}, {Tarle}, {Vikram}, \& {DES
  Collaboration}}]{martinezvazquez19}
{Mart{\'\i}nez-V{\'a}zquez}, C.~E., {Vivas}, A.~K., {Gurevich}, M., {et~al.}
  2019, \mnras, 490, 2183

\bibitem[{{McConnachie} \& {Venn}(2020)}]{mv20}
{McConnachie}, A.~W., \& {Venn}, K.~A. 2020, \aj, 160, 124

\bibitem[{{McConnachie} {et~al.}(2008){McConnachie}, {Huxor}, {Martin},
  {Irwin}, {Chapman}, {Fahlman}, {Ferguson}, {Ibata}, {Lewis}, {Richer}, \&
  {Tanvir}}]{mcconnachie08}
{McConnachie}, A.~W., {Huxor}, A., {Martin}, N.~F., {et~al.} 2008, \apj, 688,
  1009

\bibitem[{{McCrea}(1964)}]{mccrea64}
{McCrea}, W.~H. 1964, \mnras, 128, 147

\bibitem[{{Meadows} {et~al.}(2020){Meadows}, {Navarro}, {Santos-Santos},
  {Ben{\'\i}tez-Llambay}, \& {Frenk}}]{meadows20}
{Meadows}, N., {Navarro}, J.~F., {Santos-Santos}, I., {Ben{\'\i}tez-Llambay},
  A., \& {Frenk}, C. 2020, \mnras, 491, 3336

\bibitem[{{Monelli} {et~al.}(2010){Monelli}, {Gallart}, {Hidalgo}, {Aparicio},
  {Skillman}, {Cole}, {Weisz}, {Mayer}, {Bernard}, {Cassisi}, {Dolphin},
  {Drozdovsky}, \& {Stetson}}]{monelli10b}
{Monelli}, M., {Gallart}, C., {Hidalgo}, S.~L., {et~al.} 2010, \apj, 722, 1864

\bibitem[{{Moore}(1994)}]{moore94}
{Moore}, B. 1994, \nat, 370, 629

\bibitem[{{Mu{\~n}oz} {et~al.}(2018){Mu{\~n}oz}, {C{\^o}t{\'e}}, {Santana},
  {Geha}, {Simon}, {Oyarz{\'u}n}, {Stetson}, \& {Djorgovski}}]{munoz18}
{Mu{\~n}oz}, R.~R., {C{\^o}t{\'e}}, P., {Santana}, F.~A., {et~al.} 2018, \apj,
  860, 66

\bibitem[{{Mu{\~n}oz} {et~al.}(2012){Mu{\~n}oz}, {Padmanabhan}, \&
  {Geha}}]{munoz12}
{Mu{\~n}oz}, R.~R., {Padmanabhan}, N., \& {Geha}, M. 2012, \apj, 745, 127

\bibitem[{{Musella} {et~al.}(2018){Musella}, {Di Criscienzo}, {Marconi},
  {Raimondo}, {Ripepi}, {Cignoni}, {Bono}, {Brocato}, {Dall'Ora}, {Ferraro},
  {Grado}, {Iannicola}, {Limatola}, {Molinaro}, {Moretti}, {Stetson},
  {Capaccioli}, {Cioni}, {Getman}, \& {Schipani}}]{musella18}
{Musella}, I., {Di Criscienzo}, M., {Marconi}, M., {et~al.} 2018, \mnras, 473,
  3062

\bibitem[{{Mutlu-Pakdil} {et~al.}(2018){Mutlu-Pakdil}, {Sand}, {Carlin},
  {Spekkens}, {Caldwell}, {Crnojevi{\'c}}, {Hughes}, {Willman}, \&
  {Zaritsky}}]{mp18}
{Mutlu-Pakdil}, B., {Sand}, D.~J., {Carlin}, J.~L., {et~al.} 2018, \apj, 863,
  25

\bibitem[{{Mutlu-Pakdil} {et~al.}(2019){Mutlu-Pakdil}, {Sand}, {Walker},
  {Caldwell}, {Carlin}, {Collins}, {Crnojevi{\'c}}, {Mateo}, {Olszewski},
  {Seth}, {Strader}, {Willman}, \& {Zaritsky}}]{mp19}
{Mutlu-Pakdil}, B., {Sand}, D.~J., {Walker}, M.~G., {et~al.} 2019, \apj, 885,
  53

\bibitem[{{Nadler} {et~al.}(2020){Nadler}, {Drlica-Wagner}, {Bechtol}, {Mau},
  {Wechsler}, {Gluscevic}, {Boddy}, {Pace}, {Li}, {McNanna}, {Riley},
  {Garc{\'\i}a-Bellido}, {Mao}, {Green}, {Burke}, {Peter}, {Jain}, {Abbott},
  {Aguena}, {Allam}, {Annis}, {Avila}, {Brooks}, {Carrasco Kind}, {Carretero},
  {Costanzi}, {da Costa}, {De Vicente}, {Desai}, {Diehl}, {Doel}, {Everett},
  {Evrard}, {Flaugher}, {Frieman}, {Gerdes}, {Gruen}, {Gruendl}, {Gschwend},
  {Gutierrez}, {Hinton}, {Honscheid}, {Huterer}, {James}, {Krause}, {Kuehn},
  {Kuropatkin}, {Lahav}, {Maia}, {Marshall}, {Menanteau}, {Miquel}, {Palmese},
  {Paz-Chinch{\'o}n}, {Plazas}, {Romer}, {Sanchez}, {Scarpine}, {Serrano},
  {Sevilla-Noarbe}, {Smith}, {Soares-Santos}, {Suchyta}, {Swanson}, {Tarle},
  {Tucker}, {Walker}, \& {Wester}}]{Nadler20}
{Nadler}, E.~O., {Drlica-Wagner}, A., {Bechtol}, K., {et~al.} 2020, arXiv
  e-prints, arXiv:2008.00022

\bibitem[{{Ness} {et~al.}(2015){Ness}, {Hogg}, {Rix}, {Ho}, \&
  {Zasowski}}]{ness15}
{Ness}, M., {Hogg}, D.~W., {Rix}, H.~W., {Ho}, A. Y.~Q., \& {Zasowski}, G.
  2015, \apj, 808, 16

\bibitem[{{Nishikawa} {et~al.}(2020){Nishikawa}, {Boddy}, \&
  {Kaplinghat}}]{nishikawa20}
{Nishikawa}, H., {Boddy}, K.~K., \& {Kaplinghat}, M. 2020, \prd, 101, 063009

\bibitem[{{Oguri} {et~al.}(2018){Oguri}, {Diego}, {Kaiser}, {Kelly}, \&
  {Broadhurst}}]{oguri18}
{Oguri}, M., {Diego}, J.~M., {Kaiser}, N., {Kelly}, P.~L., \& {Broadhurst}, T.
  2018, \prd, 97, 023518

\bibitem[{{Pace} \& {Li}(2019)}]{pl19}
{Pace}, A.~B., \& {Li}, T.~S. 2019, \apj, 875, 77

\bibitem[{{Pace} {et~al.}(2020){Pace}, {Kaplinghat}, {Kirby}, {Simon},
  {Tollerud}, {Mu{\~n}oz}, {C{\^o}t{\'e}}, {Djorgovski}, \& {Geha}}]{pace20}
{Pace}, A.~B., {Kaplinghat}, M., {Kirby}, E., {et~al.} 2020, \mnras, 495, 3022

\bibitem[{{Paust} {et~al.}(2007){Paust}, {Chaboyer}, \& {Sarajedini}}]{paust07}
{Paust}, N. E.~Q., {Chaboyer}, B., \& {Sarajedini}, A. 2007, \aj, 133, 2787

\bibitem[{{Plummer}(1911)}]{plummer11}
{Plummer}, H.~C. 1911, \mnras, 71, 460

\bibitem[{{Portegies Zwart} {et~al.}(1997){Portegies Zwart}, {Hut}, {McMillan},
  \& {Verbunt}}]{pz97}
{Portegies Zwart}, S.~F., {Hut}, P., {McMillan}, S. L.~W., \& {Verbunt}, F.
  1997, \aap, 328, 143

\bibitem[{{Preston} \& {Sneden}(2000)}]{ps00}
{Preston}, G.~W., \& {Sneden}, C. 2000, \aj, 120, 1014

\bibitem[{{Read} {et~al.}(2006){Read}, {Goerdt}, {Moore}, {Pontzen}, {Stadel},
  \& {Lake}}]{read06}
{Read}, J.~I., {Goerdt}, T., {Moore}, B., {et~al.} 2006, \mnras, 373, 1451

\bibitem[{{Relatores} {et~al.}(2019){Relatores}, {Newman}, {Simon}, {Ellis},
  {Truong}, {Blitz}, {Bolatto}, {Martin}, {Matuszewski}, {Morrissey}, \&
  {Neill}}]{relatores19}
{Relatores}, N.~C., {Newman}, A.~B., {Simon}, J.~D., {et~al.} 2019, \apj, 887,
  94

\bibitem[{{Richardson} {et~al.}(2011){Richardson}, {Irwin}, {McConnachie},
  {Martin}, {Dotter}, {Ferguson}, {Ibata}, {Chapman}, {Lewis}, {Tanvir}, \&
  {Rich}}]{richardson11}
{Richardson}, J.~C., {Irwin}, M.~J., {McConnachie}, A.~W., {et~al.} 2011, \apj,
  732, 76

\bibitem[{{Ricotti} \& {Gnedin}(2005)}]{rg05}
{Ricotti}, M., \& {Gnedin}, N.~Y. 2005, \apj, 629, 259

\bibitem[{{Robin} {et~al.}(2003){Robin}, {Reyl{\'e}}, {Derri{\`e}re}, \&
  {Picaud}}]{besancon}
{Robin}, A.~C., {Reyl{\'e}}, C., {Derri{\`e}re}, S., \& {Picaud}, S. 2003,
  \aap, 409, 523

\bibitem[{{Rodriguez Wimberly} {et~al.}(2019){Rodriguez Wimberly}, {Cooper},
  {Fillingham}, {Boylan-Kolchin}, {Bullock}, \& {Garrison-Kimmel}}]{rw19}
{Rodriguez Wimberly}, M.~K., {Cooper}, M.~C., {Fillingham}, S.~P., {et~al.}
  2019, \mnras, 483, 4031

\bibitem[{{Rozyczka} {et~al.}(2012){Rozyczka}, {Kaluzny}, {Pietrukowicz},
  {Pych}, {Catelan}, \& {Contreras}}]{rozyczka12}
{Rozyczka}, M., {Kaluzny}, J., {Pietrukowicz}, P., {et~al.} 2012, \aap, 537,
  A89

\bibitem[{{Ryan-Weber} {et~al.}(2008){Ryan-Weber}, {Begum}, {Oosterloo}, {Pal},
  {Irwin}, {Belokurov}, {Evans}, \& {Zucker}}]{ryanweber08}
{Ryan-Weber}, E.~V., {Begum}, A., {Oosterloo}, T., {et~al.} 2008, \mnras, 384,
  535

\bibitem[{{Sales Silva} {et~al.}(2014){Sales Silva}, {Pe{\~n}a Su{\'a}rez},
  {Katime Santrich}, {Pereira}, {Drake}, \& {Roig}}]{salessilva14}
{Sales Silva}, J.~V., {Pe{\~n}a Su{\'a}rez}, V.~J., {Katime Santrich}, O.~J.,
  {et~al.} 2014, \aj, 148, 83

\bibitem[{{Santana} {et~al.}(2013){Santana}, {Mu{\~n}oz}, {Geha},
  {C{\^o}t{\'e}}, {Stetson}, {Simon}, \& {Djorgovski}}]{santana13}
{Santana}, F.~A., {Mu{\~n}oz}, R.~R., {Geha}, M., {et~al.} 2013, \apj, 774, 106

\bibitem[{{Schlafly} \& {Finkbeiner}(2011)}]{sf11}
{Schlafly}, E.~F., \& {Finkbeiner}, D.~P. 2011, \apj, 737, 103

\bibitem[{{Schlegel} {et~al.}(1998){Schlegel}, {Finkbeiner}, \&
  {Davis}}]{sfd98}
{Schlegel}, D.~J., {Finkbeiner}, D.~P., \& {Davis}, M. 1998, \apj, 500, 525

\bibitem[{{Schutz}(2020)}]{schutz20}
{Schutz}, K. 2020, \prd, 101, 123026

\bibitem[{{S{\'e}rsic}(1963)}]{sersic63}
{S{\'e}rsic}, J.~L. 1963, Boletin de la Asociacion Argentina de Astronomia La
  Plata Argentina, 6, 41

\bibitem[{{Skillman} {et~al.}(2017){Skillman}, {Monelli}, {Weisz}, {Hidalgo},
  {Aparicio}, {Bernard}, {Boylan-Kolchin}, {Cassisi}, {Cole}, {Dolphin},
  {Ferguson}, {Gallart}, {Irwin}, {Martin}, {Mart{\'\i}nez-V{\'a}zquez},
  {Mayer}, {McConnachie}, {McQuinn}, {Navarro}, \& {Stetson}}]{skillman17}
{Skillman}, E.~D., {Monelli}, M., {Weisz}, D.~R., {et~al.} 2017, \apj, 837, 102

\bibitem[{{Sollima} {et~al.}(2006){Sollima}, {Cacciari}, \&
  {Valenti}}]{sollima06}
{Sollima}, A., {Cacciari}, C., \& {Valenti}, E. 2006, \mnras, 372, 1675

\bibitem[{{Spekkens} {et~al.}(2014){Spekkens}, {Urbancic}, {Mason}, {Willman},
  \& {Aguirre}}]{spekkens14}
{Spekkens}, K., {Urbancic}, N., {Mason}, B.~S., {Willman}, B., \& {Aguirre},
  J.~E. 2014, \apjl, 795, L5

\bibitem[{{Stetson}(1987)}]{stetson87}
{Stetson}, P.~B. 1987, \pasp, 99, 191

\bibitem[{{Stryker}(1993)}]{stryker93}
{Stryker}, L.~L. 1993, \pasp, 105, 1081

\bibitem[{{Teyssier} {et~al.}(2012){Teyssier}, {Johnston}, \&
  {Kuhlen}}]{teyssier12}
{Teyssier}, M., {Johnston}, K.~V., \& {Kuhlen}, M. 2012, \mnras, 426, 1808

\bibitem[{{van den Bergh}(1999)}]{vdb99}
{van den Bergh}, S. 1999, \aapr, 9, 273

\bibitem[{Van Der~Walt {et~al.}(2011)Van Der~Walt, Colbert, \&
  Varoquaux}]{van2011numpy}
Van Der~Walt, S., Colbert, S.~C., \& Varoquaux, G. 2011, Computing in Science
  \& Engineering, 13, 22

\bibitem[{{VandenBerg} {et~al.}(2014){VandenBerg}, {Bergbusch}, {Ferguson}, \&
  {Edvardsson}}]{vdb14}
{VandenBerg}, D.~A., {Bergbusch}, P.~A., {Ferguson}, J.~W., \& {Edvardsson}, B.
  2014, \apj, 794, 72

\bibitem[{{Vandenberg} {et~al.}(1990){Vandenberg}, {Bolte}, \&
  {Stetson}}]{vdb90}
{Vandenberg}, D.~A., {Bolte}, M., \& {Stetson}, P.~B. 1990, \aj, 100, 445

\bibitem[{{Vivas} {et~al.}(2020){Vivas}, {Mart{\'\i}nez-V{\'a}zquez}, \&
  {Walker}}]{vivas20}
{Vivas}, A.~K., {Mart{\'\i}nez-V{\'a}zquez}, C., \& {Walker}, A.~R. 2020,
  \apjs, 247, 35

\bibitem[{{Walker} {et~al.}(2007){Walker}, {Mateo}, {Olszewski}, {Gnedin},
  {Wang}, {Sen}, \& {Woodroofe}}]{walker07}
{Walker}, M.~G., {Mateo}, M., {Olszewski}, E.~W., {et~al.} 2007, \apjl, 667,
  L53

\bibitem[{{Wang} {et~al.}(2019){Wang}, {Koposov}, {Drlica-Wagner}, {Pieres},
  {Li}, {de Boer}, {Bechtol}, {Belokurov}, {Pace}, {Bacon}, {Abbott}, {Annis},
  {Bertin}, {Brooks}, {Buckley-Geer}, {Burke}, {Carnero Rosell}, {Carrasco
  Kind}, {Carretero}, {da Costa}, {De Vicente}, {Desai}, {Diehl}, {Doel},
  {Estrada}, {Flaugher}, {Fosalba}, {Frieman}, {Garc{\'\i}a-Bellido}, {Gerdes},
  {Gruen}, {Gruendl}, {Gschwend}, {Gutierrez}, {Hollowood}, {Honscheid},
  {Hoyle}, {James}, {Kent}, {Kuehn}, {Kuropatkin}, {Maia}, {Marshall},
  {Menanteau}, {Miquel}, {Plazas}, {Sanchez}, {Santiago}, {Scarpine},
  {Schindler}, {Schubnell}, {Serrano}, {Sevilla-Noarbe}, {Smith}, {Smith},
  {Sobreira}, {Suchyta}, {Swanson}, {Tarle}, {Thomas}, {Tucker}, {Walker}, \&
  {DES Collaboration}}]{wang19}
{Wang}, M.~Y., {Koposov}, S., {Drlica-Wagner}, A., {et~al.} 2019, \apjl, 875,
  L13

\bibitem[{{Warnick} {et~al.}(2008){Warnick}, {Knebe}, \& {Power}}]{warnick08}
{Warnick}, K., {Knebe}, A., \& {Power}, C. 2008, \mnras, 385, 1859

\bibitem[{{Watkins} {et~al.}(2019){Watkins}, {van der Marel}, {Sohn}, \&
  {Evans}}]{watkins19}
{Watkins}, L.~L., {van der Marel}, R.~P., {Sohn}, S.~T., \& {Evans}, N.~W.
  2019, \apj, 873, 118

\bibitem[{{Weisz} {et~al.}(2014){Weisz}, {Dolphin}, {Skillman}, {Holtzman},
  {Gilbert}, {Dalcanton}, \& {Williams}}]{weisz14a}
{Weisz}, D.~R., {Dolphin}, A.~E., {Skillman}, E.~D., {et~al.} 2014, \apj, 789,
  147

\bibitem[{{Weisz} {et~al.}(2012){Weisz}, {Zucker}, {Dolphin}, {Martin}, {de
  Jong}, {Holtzman}, {Dalcanton}, {Gilbert}, {Williams}, {Bell}, {Belokurov},
  \& {Evans}}]{weisz12}
{Weisz}, D.~R., {Zucker}, D.~B., {Dolphin}, A.~E., {et~al.} 2012, \apj, 748, 88

\bibitem[{{Westmeier} {et~al.}(2015){Westmeier}, {Staveley-Smith},
  {Calabretta}, {Jurek}, {Koribalski}, {Meyer}, {Popping}, \&
  {Wong}}]{westmeier15}
{Westmeier}, T., {Staveley-Smith}, L., {Calabretta}, M., {et~al.} 2015, \mnras,
  453, 338

\bibitem[{{Wheeler} {et~al.}(2019){Wheeler}, {Hopkins}, {Pace},
  {Garrison-Kimmel}, {Boylan-Kolchin}, {Wetzel}, {Bullock}, {Kere{\v{s}}},
  {Faucher-Gigu{\`e}re}, \& {Quataert}}]{wheeler19}
{Wheeler}, C., {Hopkins}, P.~F., {Pace}, A.~B., {et~al.} 2019, \mnras, 490,
  4447

\bibitem[{{White} \& {Shawl}(1987)}]{ws87}
{White}, R.~E., \& {Shawl}, S.~J. 1987, \apj, 317, 246

\bibitem[{{Willman} {et~al.}(2005){Willman}, {Blanton}, {West}, {Dalcanton},
  {Hogg}, {Schneider}, {Wherry}, {Yanny}, \& {Brinkmann}}]{willman05a}
{Willman}, B., {Blanton}, M.~R., {West}, A.~A., {et~al.} 2005, \aj, 129, 2692

\bibitem[{{Zoutendijk} {et~al.}(2020){Zoutendijk}, {Brinchmann}, {Boogaard},
  {Gunawardhana}, {Husser}, {Kamann}, {Ramos Padilla}, {Roth}, {Bacon}, {den
  Brok}, {Dreizler}, \& {Krajnovi{\'c}}}]{zoutendijk20}
{Zoutendijk}, S.~L., {Brinchmann}, J., {Boogaard}, L.~A., {et~al.} 2020, \aap,
  635, A107

\bibitem[{{Zucker} {et~al.}(2006){Zucker}, {Belokurov}, {Evans}, {Wilkinson},
  {Irwin}, {Sivarani}, {Hodgkin}, {Bramich}, {Irwin}, {Gilmore}, {Willman},
  {Vidrih}, {Fellhauer}, {Hewett}, {Beers}, {Bell}, {Grebel}, {Schneider},
  {Newberg}, {Wyse}, {Rockosi}, {Yanny}, {Lupton}, {Smith}, {Barentine},
  {Brewington}, {Brinkmann}, {Harvanek}, {Kleinman}, {Krzesinski}, {Long},
  {Nitta}, \& {Snedden}}]{zucker06b}
{Zucker}, D.~B., {Belokurov}, V., {Evans}, N.~W., {et~al.} 2006, \apjl, 643,
  L103

\bibitem[{{Zumalac{\'a}rregui} \& {Seljak}(2018)}]{zs18}
{Zumalac{\'a}rregui}, M., \& {Seljak}, U. 2018, \prl, 121, 141101

\end{thebibliography}

\clearpage

\end{document}